\documentclass[
 reprint,
 amsmath,amssymb,
 aps,
]{revtex4-2}

\usepackage{bm}
\usepackage{dcolumn}
\usepackage{graphicx}
\usepackage{mathtools}
\usepackage[T1]{fontenc}

\newcommand{\varA}[1]{{\operatorname{#1}}}

\begin{document}

\preprint{APS/123-QED}

\title{Ergodic characterization of non-ergodic anomalous diffusion processes}

\author{Madhur Mangalam}
\email{Correspondence should be sent to: \href{mailto:mmangalam@uomaha.edu}{mmangalam@uomaha.edu}.}
\affiliation{Division of Biomechanics and Research Development, Department of Biomechanics, and Center for Research in Human Movement Variability, University of Nebraska at Omaha, Omaha, NE 68182, USA}

\author{Ralf Metzler}
\affiliation{Institute of Physics \& Astronomy, University of Potsdam, 14776 Potsdam, Germany}

\author{Damian G. Kelty-Stephen}
\affiliation{Department of Psychology, State University of New York at New Paltz, New Paltz, New York 12561, USA}

\begin{abstract}
Anomalous diffusion in a variety of complex systems abounds in nature and spans multiple space and time scales. Canonical characterization techniques that rely upon mean squared displacement ($\varA{MSD}$) break down for non-ergodic processes, making it challenging to characterize anomalous diffusion from an individual time-series measurement. Non-ergodicity reigns when the time-averaged mean square displacement $\varA{TA-MSD}$ differs from the ensemble-averaged mean squared displacement $\varA{EA-MSD}$ even in the limit of long measurement series. In these cases, the typical theoretical results for ensemble averages cannot be used to understand and interpret data acquired from time averages. The difficulty then lies in obtaining statistical descriptors of the measured diffusion process that are not non-ergodic. We show that linear descriptors such as the standard deviation ($SD$), coefficient of variation ($CV$), and root mean square ($RMS$) break ergodicity in proportion to non-ergodicity in the diffusion process. In contrast, time series of descriptors addressing sequential structure and its potential nonlinearity: multifractality change in a time-independent way and fulfill the ergodic assumption, largely independent of the time series’ non-ergodicity. We show that these findings follow the multiplicative cascades underlying these diffusion processes. Adding fractal and multifractal descriptors to typical linear descriptors would improve the characterization of anomalous diffusion processes. Two particular points bear emphasis here. First, as an appropriate formalism for encoding the nonlinearity that might generate non-ergodicity, multifractal modeling offers descriptors that can behave ergodically enough to meet the needs of linear modeling. Second, this capacity to describe non-ergodic processes in ergodic terms offers the possibility that multifractal modeling could unify several disparate non-ergodic diffusion processes into a common framework.
\end{abstract}


\maketitle

\section{Introduction}

\begin{figure*}[t]
\includegraphics{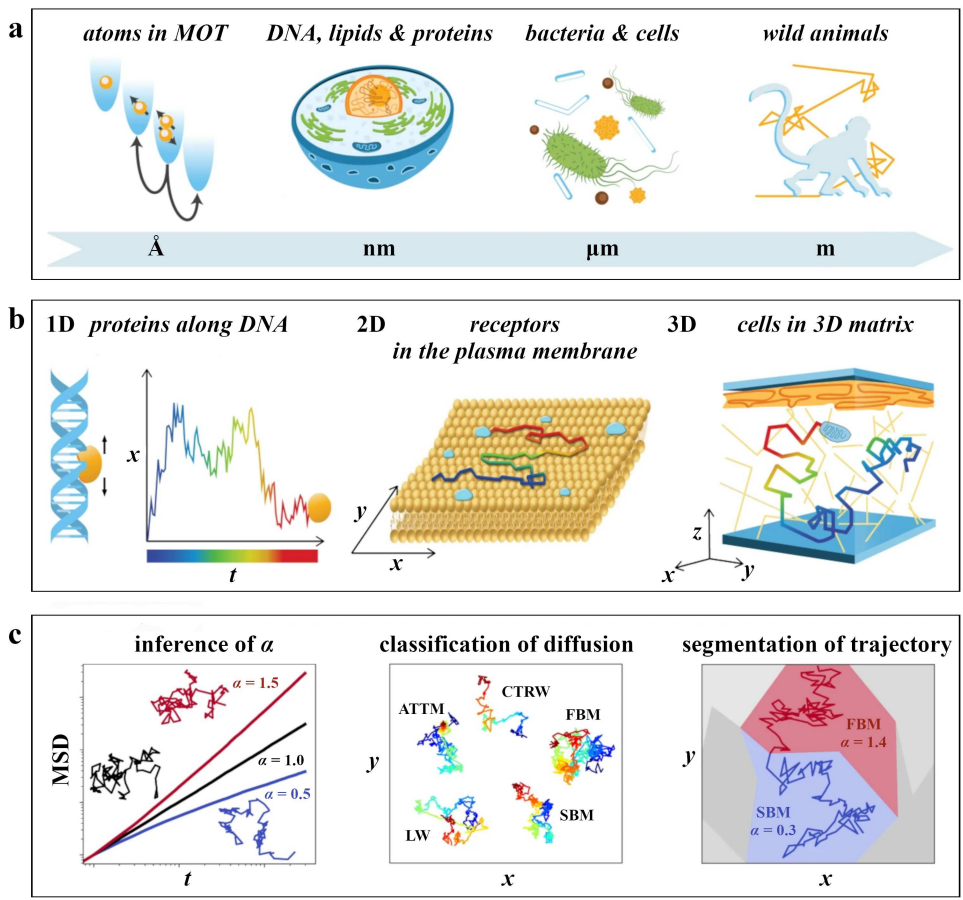}
\caption{Random walks or anomalous diffusion processes, which are defined by unpredictable variations in an observable, can be found in a wide range of systems over a wide range of spatial and temporal scales. (a) Examples include atoms in magneto-optical traps, the diffusion of biological components such as DNA, lipids, and proteins, bacteria and cell motility, and foraging wild animals. (b) Random walks in three-dimensional spaces may be dimensionally reduced: 1D, proteins sliding along DNA segments; 2D, receptors diffusing across the plasma membrane; 3D, cells migrating through a three-dimensional matrix. The color code of the trajectories represents time. (c) Representative trajectories and corresponding $\mathrm{MSD}$ for diffusive ($\alpha=1$, black lines), subdiffusive ($0<\alpha<1$, blue lines), and superdiffusive ($1<\alpha<2$, red lines) motion. The underlying anomalous diffusion model can be classified as fractional Brownian motion (FBM), scaled Brownian motion (SBM), continuous-time random walk (CTRW), annealed transient time motion (ATTM), or Lévy walk (LW). These diffusion models produce subtle changes (details of the models are described in “Methods” section). A trajectory can show a change-point by switching the diffusion model or exponent as a result of diffusion in a spatially heterogeneous environment. Adapted from Muñoz-Gil et al. \cite{munoz2021objective}.}
\label{fig:f1}
\end{figure*}

Anomalous diffusion abounds in nature---atoms in magneto-optical traps \cite{sagi2012observation,zhao2014direct}, DNA, lipids, and proteins \cite{banks2005anomalous,barkai2012single,guigas2008sampling,hofling2013anomalous,jeon2011vivo,jeon2012anomalous,krapf2019strange,metzler2000random,ritchie2005detection,tolic2004anomalous}, bacteria, cells, and parasites \cite{angelini2011glass,dieterich2008anomalous,dieterich2022anomalous,golding2006physical,hapca2009anomalous,lagarde2020colloidal,mukherjee2021anomalous}, foraging wild animals \cite{benhamou2007many,james2011assessing,reynolds2009levy} and human hunter gatherers \cite{brown2007levy,raichlen2014evidence}, economic markets \cite{cherstvy2021scaled,plerou2000economic,vazquez2006modeling}, and various other processes \cite{oliveira2019anomalous,sokolov2005diffusion,timashev2010anomalous,vilk2022unravelling} show anomalous diffusion that spans multiple scales (Fig.~\ref{fig:f1}a). All these processes are characterized by an erratic change of an observable (e.g., position, temperature, or stock price) over time (Fig.~\ref{fig:f1}b). ``Anomalous” implies that the observable $x$’s mean squared displacement ($\varA{MSD}$) does not grow linearly with time $t$, $\langle x^{2}
(t) \rangle \propto t$, as predicted by Fick’s theory of diffusion, but follows another power–law pattern $\langle x^{2}(t) \rangle \propto t^{\alpha}$, with $\alpha \neq 1$. Frequently, $\alpha < 1$, indicating subdiffusion \cite{bancaud2009molecular,caspi2000enhanced,golding2006physical,hofling2011anomalous,seisenberger2001real,smith1999anomalous,weber2010bacterial,weber2010bacterial,weber2010subdiffusive}. Superdiffusion---characterized by $\alpha > 1$---is less commonly reported than subdiffusion but is often observed in active physical and biological systems \cite{arcizet2008temporal,caspi2000enhanced,caspi2002diffusion,de2011levy,duits2009mapping,gonzalez2008understanding,leoni2014structural,mashanova2010evidence,nathan2008movement}.

Widespread evidence of anomalous diffusion processes has sparked a major theoretical effort to comprehend and formally organize the mechanisms that might explain them. This endeavor has resulted in various mathematical models with and without long-range correlations and varied spatial (step length) and temporal (step duration) random distributions. Models of anomalous diffusion have grown from elaborations of Brownian motion, which depicts the movement of a small particle in a fluid due to thermal forces. Brownian motion embodies ordinary diffusion when $\mathrm{MSD}$ grows linearly with time, $\langle x^{2}(t) \rangle \propto t$ \cite{pearson1905problem}. In a curious turn of terminology, anomalous diffusion emerges as the more general case---``anomalous is normal" \cite{klafter2005anomalous}, and these widespread applications of diffusion modeling come from adding anomalous nuance to these fluctuation patterns in the narrow case of so-called ``ordinary” diffusion \cite{klafter2005anomalous}. The anomalous diffusion underlying an observed process can be modeled as fractional Brownian motion (FBM) \cite{mandelbrot1968fractional}, scaled Brownian motion (SBM) \cite{lim2002self,jeon2014scaled}, noisy continuous-time random walk (nCTRW) \cite{jeon2013noisy}---a variant of CTRW \cite{scher1975anomalous}, annealed transient time motion (ATTM) \cite{massignan2014nonergodic}, or Lévy walk (LW) \cite{klafter1994levy}. These diffusion processes show subtle differences in how fluctuations are distributed in time (for details of the models, see Appendix A). The theoretical challenge for explaining anomalous diffusion compounds with the observation that these different processes are not mutually exclusive---they may only reflect distinct modes into which the same observable system can transition \cite{lanoiselee2019non,munoz2021objective}. A single observed trajectory can switch at change points among regimes best explained by different of these models as a result of interactions with the surroundings in a heterogeneous environment \cite{banks2005anomalous,banks2016characterizing,dix2008crowding,jeon2011vivo,jeon2012anomalous,molina2018crossover,sokolov2012models} (Fig.~\ref{fig:f1}c). Often, the observed motion also simultaneously exhibits features of more than a single of the ``elementary" stochastic models. Despite their common heritage in Brownian motion \cite{abe2005anomalous}, these processes exhibit seemingly disparate modes of nonlinear and non-ergodic behavior \cite{klafter2005anomalous,klafter2011first,metzler2004restaurant,sancho2004diffusion,sokolov2002fractional,sokolov2012models}. The challenge is that modeling the causal evolution from one mode to another requires causal models, and prevailing statistical approaches to causal modeling largely assume linearity and ergodicity.

A leading motivation to study anomalous diffusion models is to detect and classify specific anomalous diffusion processes in empirical data. However, all the above factors make this classification a challenging feat. Therefore, recent attempts include Bayesian \cite{krog2017bayesian,krog2018bayesian,park2021bayesian,thapa2018bayesian,thapa2022bayesian} as well as machine learning (ML) approaches \cite{bo2019measurement,cichos2020machine,granik2019single,janczura2020classification,munoz2020single,munoz2021objective}, and even unsupervised approaches \cite{pineda2022geometric,gajowczyk2021detection,gentili2021characterization,kowalek2022boosting,munoz2021unsupervised,seckler2022bayesian}. However, these attempts are based on predominantly atheoretical selection of features which may not necessarily be related to plausible generating mechanisms \cite{kowalek2019classification,loch2020impact}. A more theoretically defined set of features can potentially improve the ML-powered characterization of anomalous diffusion processes in empirical data.

One possible resolution lies in the observation that temporal correlations and non-Gaussianity are common features of multifractal processes. Multifractal geometry is also a formalism that specifically addresses the intermittent, non-ergodic fluctuations across a wide range of scales and the nonlinear interactions of short-range events with large-scale contextual factors \cite{mandelbrot1974intermittent,schertzer1997multifractal}. Multifractality is observed in strong anomalous diffusion \cite{gal2010experimental,rebenshtok2007distribution}. Thus, these anomalous processes with common heritage in the Brownian-motion formalism may find a common reunified description. This point is not to say that the models generating these different regimes of anomalous diffusion are explicitly multifractal. Instead, it is to recognize that multifractal geometry has long been considered a modeling framework broad enough to explain how these different modes of anomalous diffusion evolve over time and sometimes with change points from one mode to another \cite{shlesinger1987levy}. Critically, the first step towards explanation through prevailing causal models is meeting the basic benchmark of ergodicity for the traditionally linear statistical structure of causal modeling. We use numerical simulations to test the hypothesis that multifractal geometrical estimates of these diffusive properties offer an ergodic descriptor that makes these disparate diffusion processes amenable to linear causal framework.

Efforts so far have addressed the twofold challenges of properly quantifying empirical diffusion processes and of doing so with the appropriate model. The difficulty here is that both concerns must be pursued largely in tandem: we must empirically estimate the value of model parameters, and to ensure these estimates are effective, we must be sure to use a model appropriate to the data. Best practices for balancing parameter estimation with model specification includes analyzing the empirical time-series data with various statistical observables such as $\varA{MSD}$, spectral power analysis, Van-Hove correlation functions, step-length or flight-time distribution, and ergodicity breaking parameter \cite{ernst2014probing,he2008random,kepten2015guidelines,krapf2019spectral,magdziarz2009fractional,rytov1989principles,slkezak2019codifference,sposini2020universal,thirumalai1989ergodic,vilk2022classification,vilk2022unravelling}. However, data interpretation can be subjective and is contingent on the fidelity of the observed data which is inevitably constrained by length and number of observations, measurement noise, and sample spatiotemporal heterogeneity.

It is important to note that the preceding best practices are sometimes at odds with the characteristic non-ergodicity of many anomalous-diffusion processes. That is to say, the constraints imposed by ergodicity on diffusion modeling set in well before any thoughts about linear causal models to articulate any causal developmental relationships among disparate anomalous-diffusion processes. Non-ergodicity entails a failure of individual time-series measurements to represent an ensemble. Non-ergodicity reigns when $\varA{TA-MSD}$ differs from $\varA{EA-MSD}$. Sample-size constraints on measurements of non-ergodic processes thus dramatically constrain the interpretation of the canonical characterization techniques that rely on $\varA{MSD}$. In these cases, the typical theoretical results for ensemble averages cannot be used to understand and interpret data acquired from time averages. For example, FBM is ergodic for $\alpha = 0.1$, though convergence of the $\varA{EA-MSD}$ to $\varA{TA-MSD}$ may be slower for values of the anomalous exponent close to 1 \cite{deng2009ergodic}. The ergodicity in FBM requires careful analysis as a function of $\alpha$ \cite{kelty2022fractal,kelty2023multifractal,mangalam2021point,mangalam2022ergodic}, and often higher order moments accounting for the skewness and kurtosis are necessary to study ergodicity breaking in FBM \cite{schwarzl2017quantifying}. CTRW, ATTM, and SBM show weak ergodicity breaking \cite{bel2005weak,burov2011single,cherstvy2014nonergodicity,jeon2011vivo,massignan2014nonergodic,rebenshtok2007distribution,safdari2015quantifying}. Finally, LW shows a distinct kind of ergodicity breaking---named ultra-weak non-ergodicity---in which ensemble and time averages only differ by a constant factor \cite{froemberg2013time,godec2013finite}. Deriving inferences from canonical estimates submitted to linear causal models makes the questionable compromise of enforcing similarity while neglecting diversity for formal convenience, given the variability in the ergodic features of these diffusion processes. Moreover, such inferences may obscure any artifacts of non-ergodicity or fail to articulate the systematic changes that lead to non-ergodicity, potentially obscuring any genuine individual differences and discarding any generalizable truths we might have gleaned from the same diversity that was intended to represent these disparate models of nonlinearity and non-ergodicity \cite{mangalam2021point}.

Here, we have used cascade-dynamical descriptors rooted in the multifractal formalism to compare the ergodicity-related diffusive properties of various anomalous processes. An important theoretical move beyond attempts at formal convenience may be explicitly addressing the underlying mechanisms generating non-ergodicity in empirical examples of anomalous diffusion. In this sense, multifractal geometry is not merely convenient because it affords an analytical repertoire for addressing features sometimes seen in different anomalous-diffusion processes. Rather, multifractal geometry is a theoretically valid means to estimate parameters of cascade dynamics that can generate a wide variety of intermittent, non-ergodic behavior. If we encode those aspects of the diffusion process known to generate non-ergodicity, these parameter estimates might be ergodic---and indeed, current evidence shows that they are \cite{kelty2022fractal,mangalam2022ergodic}. We know, for instance, that $fGn$ observed in biological and psychological phenomena break ergodicity primarily due to the interdependencies among factors unfolding at multiple spatial and temporal scales \cite{dixon2012multifractal,ihlen2010interaction,kelty2021multifractal,kloos2010voluntary,van2003self}. The scale-invariant shape of the power-law autocorrelation in $fGn$—quantified as the fractal exponent $H_{fGn}$ shows none of the ergodicity breaking of the $fGn$ series \cite{mangalam2022ergodic,mangalam2022ergodic}. Although $fGn$ is a linear process, one possible explanation for power-law scaling is the nonlinear interactions across scales in cascade processes known to generate intermittent, non-ergodic behavior \cite{mandelbrot1974intermittent}. The strength of cascade dynamics can be quantified as the multifractal spectrum width $\Delta\alpha$ and then as \textit{t}-statistic comparing that multifractal-spectrum width to spectrum widths for linear surrogates $t_{MF}$ \cite{kelty2022multifractaltest,mangalam2020multifractal}. If cascades can explain the non-ergodic behavior of all anomalous diffusion processes, then $\Delta\alpha$ and $t_{MF}$ might avoid displaying the breaking of ergodicity and can be submitted to linear models of cause and effect. Indeed, we have shown that for $fGn$, all three descriptors: $H_{fGn}$, $\Delta\alpha$, and $t_{MF}$, avoid ergodicity breaking \cite{mangalam2022ergodic,kelty2022fractal}. Even without any interest in causal modeling to examine the developmental change among modes of anomalous diffusion, this approach of extending beyond $\mathrm{MSD}$ to more generalized multifractal modeling can help apparently restoring broken ergodicity for all forms of anomalous diffusion processes.

The structure of this article is as follows. We first provide a comparative analysis of ergodic properties of synthetic FBM, SBM, nCTRW, ATTM, and LW time series with different values of the anomalous exponent. We then provide a comparative analysis of ergodic properties of the time series of linear and cascade-dynamical descriptors of these synthetic time series. We discuss how the latter descriptors---rooted in multifractal formalism---encode the nonlinearity that might generate non-ergodicity in these anomalous diffusion processes. Finally, we discuss the implications of using cascade-dynamical descriptors of anomalous diffusion processes to meet the needs of linear modeling and classify specific anomalous diffusion processes in empirical data.

\section{Materials and Methods}

\subsection{Simulating FBM, SBM, nCTRW, ATTM, and LW series}

\begin{figure*}
\includegraphics{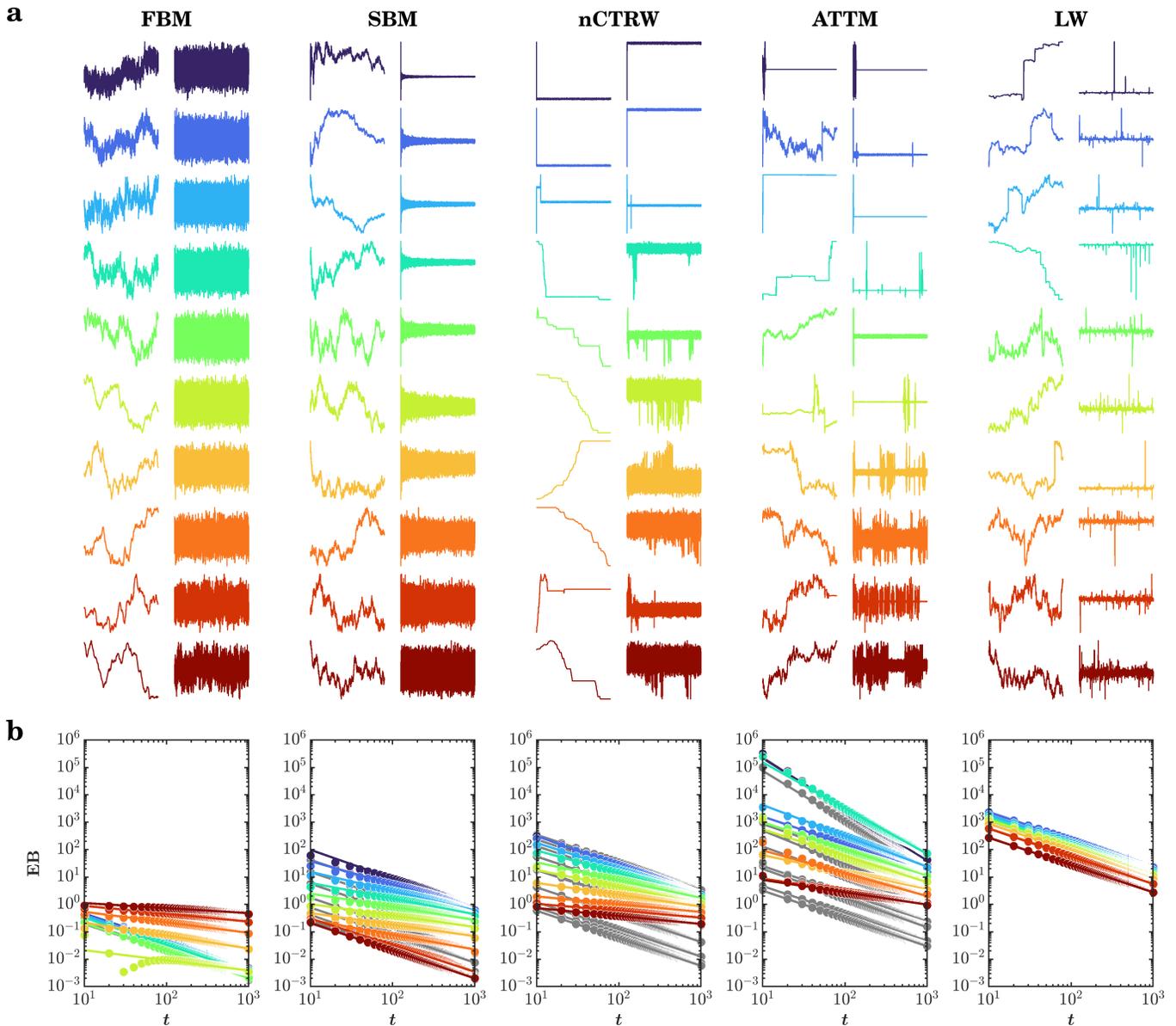}
\caption{Simulated anomalous diffusion processes and their ergodic properties. (a) Sample trajectories, and the corresponding fluctuation series, generated according to the following five models: fractional Brownian motion (FBM, ergodic), scaled Brownian motion (SBM, weakly non-ergodic), noisy continuous-time random walk (nCTRW, weakly non-ergodic), annealed transient time motion (ATTM, weakly non-ergodic), and Lévy walk (LW, ultra-weakly non-ergodic). While FBM, SBM, nCTRW, and ATTM are strictly subdiffusive ($0.1 \leq \alpha \leq 1$), LW is superdiffusive ($\alpha \geq 1$). The anomalous exponent $\alpha$ ranges from 0.1 to 1 for FBM, SBM, nCTRW, and ATTM, and from 1.1 to 2 for LW, with an increment of 0.1 from top to bottom. (b) Ergodicity-breaking parameter ($\mathrm{EB}$)-vs.-$\log t$ curves for each of the five processes ($N = 100$; lag is $\Delta = 10$ samples). The different decay rates of $\mathrm{EB}\rightarrow0$ as $t \rightarrow \infty$ for different diffusion models and exponents indicates that each process breaks ergodicity to different extents: FBM, ergodic; SBM, weakly non-ergodic; nCTRW, weakly non-ergodic; ATTM, weakly non-ergodic; and LW, ultra-weakly non-ergodic. Additionally, for SBM, nCTRW, and ATTM, $\mathrm{EB}$ differs between the original series (colored circles and lines) and their shuffled versions (grey circles and lines). For LW, $\mathrm{EB}$ for the original and shuffled series completely coincide. For FBM, $\mathrm{EB}$ for the shuffled series for all $\alpha$ coincide with $\mathrm{EB}$ for the original series with $\alpha = 0.5$. Hence, grey circles and lines are eclipsed by colored circles and lines for FBM and LW.}
\label{fig:f2}
\end{figure*}

We simulated using MATLAB (Matlab Inc, Natick, MA) 50,001-sample synthetic trajectories generated according to each of the following five different anomalous diffusion models (Fig.~\ref{fig:f2}a): (i) FBM (ergodic)---a motion with correlated long-range steps \cite{mandelbrot1968fractional}, (ii) SBM (weakly non-ergodic)---a motion whose diffusion coefficient features deterministic time-dependent changes \cite{lim2002self,jeon2014scaled}, (iii) nCTRW (weakly non-ergodic)---a variant of CTRW \cite{scher1975anomalous}, a motion undergoing local trapping with a wide distribution of waiting times \cite{jeon2013noisy}, (iv) ATTM (weakly non-ergodic)---a motion with random changes of the diffusion coefficient in time \cite{massignan2014nonergodic}, and (iv) LW (ultra-weakly non-ergodic)---a motion displaying irregular jumps with constant velocity \cite{klafter1994levy}. Appendix A describes the anomalous diffusion models considered and the algorithm used to simulate each process in more detail. The anomalous exponent was restricted to $\alpha \geq 0.1$ because smaller exponents produce practically immobile trajectories. Note that FBM, SBM, nCTRW, and ATTM are considered in the subdiffusive range $0.1 \leq \alpha \leq 1$ and LW is subdiffusive in the range $\alpha \geq 1$. We simulated 100 series using each of the 10 different exponents for each model: $\alpha = \{0.1,0.2,\dots,1\}$ for FBM, SBM, nCTRW, and ATTM, and $\alpha = \{1.1,1.2,\dots,2\}$ for LW. Each series was then differentiated and unsigned to obtain a 50,000-samples fluctuation series. All analysis was conducted on these unsigned fluctuation series because multifractal analysis require all values in a time series to be positive. Using unsigned values is a common practice in fractal and multifractal analysis. A shuffled version of each original fluctuation series was generated for comparison, because ergodicity is about how sequence exemplifies a typical mean trajectory of a sample of realizations. Shuffling breaks the sequence, producing additive white Gaussian noise (awGn) that oscillates around the mean. Finally, each original series $x(t)$ was segmented into 100 non-overlapping 500-sample segments, $s$, such that $s = \{s_{1},s_{2},\dots,s_{100}\}$. The corresponding shuffled version for each process was likewise segmented. 100-sample time series of linear and nonlinear descriptors were then obtained across these segmented series. This procedure breaks long-range correlations in the process, as we will see below.

\subsection{Estimating linear descriptors}

We computed $\varA{EA-MSD}$ and $\varA{TA-MSD}$ for each trajectory of each process. We defined $\varA{EA-MSD}$ as
\begin{equation*}
  \langle x^{2}(t) \rangle = \frac{1}{N} \sum_{i = 1}^{N} \bigl [ x_{i}(t)-x_{i}(0) \bigl ] ^2, \tag{1}\label{eq:1}
\end{equation*}
for a set of $N$ trajectories, and we defined the time average mean squared displacement ($\varA{TA-MSD}$) as
\begin{equation*}
  \overline{\delta^{2}(\tau)} = \frac{1}{L - m} \sum_{i = 1}^{L - m} \bigl [ x((i + m) \Delta t) - x(i \Delta t) \bigl ]^2, \tag{2}\label{eq:2}
\end{equation*}
when the series is sampled at $L$ discrete times $\tau = m \Delta t$.

We also computed $\mathrm{MSD}$-related three linear descriptors for each of the 100 non-overlapping 500-sample segments for the original version (i.e., unshuffled) and a shuffled version (i.e., a version with the temporal information destroyed) of each process. We defined the standard deviation ($SD$) as
\begin{equation*}
  SD = \sqrt{\frac{1}{T} \sum_{t=1}^{T} \Bigl( x(t) - \overline{x(t)} \Bigl) ^{2}}, \tag{3}\label{eq:3}
\end{equation*}
where $T$ is the fluctuation series length, and we defined the coefficient of variation ($CV$) as
\begin{equation*}
  CV = \frac{\sqrt{\frac{1}{T} \sum_{t=1}^{T} \Bigl( x(t) - \overline{x(t)} \Bigl) ^{2}}}{\overline{x(t)}}. \tag{4}\label{eq:4}
\end{equation*}
We also defined the root mean square ($RMS$), i.e.,
\begin{equation*}
  RMS = \sqrt{\frac{1}{T} \sum_{t=1}^{T} |x(t)|^{2}}. \tag{5}\label{eq:5}
\end{equation*}

\subsection{Estimating cascade-dynamical descriptors}

We computed three cascade-dynamical descriptors for each of the 100 non-overlapping 500-sample segments for the original version (i.e., unshuffled) and a shuffled version (i.e., a version with the temporal information destroyed) of each process.

\subsubsection{Accessing fractality using detrended fluctuation analysis}

Detrended fluctuation analysis (DFA) computes the Hurst exponent, $H_{fGn}$, quantifying the strength of long-range correlations in series \cite{peng1994mosaic,peng1995quantification} using the first-order integration of $T$-length time series $x(t)$:
\begin{equation*}
  y(i) = \sum_{k=1}^{i} \Bigl ( x(k) - \overline{x(t)} \Bigl ), \tag{6}\label{eq:6}
\end{equation*}
where $i = 1,2,3, \dots, T$. DFA computes root mean square ($RMS$; i.e., averaging the residuals) for each linear trend $y_{n}(t)$ fit to $N_n$ non-overlapping $n$-length bins to build a fluctuation function: 
\begin{equation*}
  f(v, n) = \sqrt{\frac{1}{N_n} \sum_{v=1}^{N_n} \biggl (\frac{1}{n} \sum_{i=1}^{n} \Bigl ( y \bigl ( (v-1)\,n+i \bigl ) - y_{v}(i) \Bigl ) ^{2} \biggl ) }, \tag{7}\label{eq:7}
\end{equation*}
where n = \{4, 8, 12, \dots \} < T/4. $f(n)$ is a power law,
\begin{equation*}
  f(n) \sim n^{H_{fGn}}, \tag{8}\label{eq:8}
\end{equation*}
where $H_{fGn}$ is the scaling exponent estimable using logarithmic transformation:
\begin{equation*}
  \log f(n) = H_{fGn}\log n. \tag{9}\label{eq:9}
\end{equation*}
Higher $H_{fGn}$ corresponds to stronger long-range correlations.

\subsubsection{Assessing multifractal spectrum width using the direct-estimation of singularity spectrum}

Chhabra and Jensen’s \cite{chhabra1989direct} direct method estimates multifractal spectrum width $\Delta\alpha$ by sampling a series $x(t)$ at progressively larger scales using the proportion of signal $P_{i}(n)$ falling within the $v$th bin of scale $n$ as
\begin{equation*}
  P_{v}(n) = \frac{\sum\limits_{k = (v-1)\,n+1}^{N_{n}} x(k)}{\sum{x(t)}},  \tag{12}\label{eq:12}
\end{equation*}
where n = \{2, 4, 8, 16, \dots \} < T/8. As $n$ increases, $P_{v}(n)$ represents a progressively larger proportion of $x(t)$,
\begin{equation*}
  P(n) \propto n^{\alpha}, \tag{11}\label{eq:11}
\end{equation*}
suggesting a growth of the proportion according to one ``singularity'' strength $\alpha$ \cite{mandelbrot1982fractal}. $P(n)$ exhibits multifractal dynamics when it grows heterogeneously across time scales $n$ according to multiple singularity strengths, such that
\begin{equation*}
  P(n_{v}) \propto n^{\alpha_{v}}, \tag{12}\label{eq:12}
\end{equation*}
whereby each $v$th bin may show a distinct relationship of $P(n)$ with $n$. The width of this singularity spectrum, $\Delta\alpha = (\alpha_{max}-\alpha_{min})$, indicates the heterogeneity of these relationships \cite{halsey1986fractal,mandelbrot2013fractals}.

Chhabra and Jensen's \cite{chhabra1989direct} method estimates $P(n)$ for $N_{n}$ non-overlapping bins of $n$-sizes and transforms them into a ``mass'' $\mu(q)$ using a $q$ parameter emphasizing higher or lower $P(n)$ for $q>1$ and $q<1$, respectively, in the form
\begin{equation*}
  \mu_{v}(q,n) = \frac{ \bigl [P_{v}(n) \bigl ] ^{q}}{\sum\limits_{j=1}^{N_{n}} \bigl[ P_{j}(n) \bigl ] ^{q}}. \tag{13}\label{eq:13}
\end{equation*}
Then, $\alpha(q)$ is the singularity for mass $\mu$-weighted $P(n)$ estimated as
\begin{equation*}
  \alpha(q) = - \lim_{N_{n}\to\infty} \frac{1}{\ln N_{n}} \sum_{v=1}^{N_{n}} \mu_{v} (q,n) \ln P_{v}(n)
\end{equation*}
\begin{equation*}
  = \lim_{n\to0} \frac{1}{\ln n} \sum_{v=1}^{N_{n}} \mu_{v} (q,n) \ln P_{v}(n). \tag{14}\label{eq:14}
\end{equation*}
Each estimated value of $\alpha(q)$ belongs to the multifractal spectrum only when the Shannon entropy of $\mu(q,n)$ scales with $n$ according to the Hausdorff dimension $f(q)$ \cite{chhabra1989direct}, where
\begin{equation*}
  f(q) = - \lim_{N_{n}\to\infty} \frac{1}{\ln N_{n}} \sum_{v=1}^{N_{n}} \mu_{v} (q,n) \ln \mu_{v}(q,n)
\end{equation*}
\begin{equation*}
  = \lim_{v\to0} \frac{1}{\ln n} \sum_{v=1}^{N_{n}} \mu_{v} (q,n) \ln \mu_{v}(q,n). \tag{15}\label{eq:15}
\end{equation*}

For values of $q$ yielding a strong relationship between Eqs.~(\ref{eq:14}) \& (\ref{eq:15})—in this study, correlation coefficient $r > 0.995$, the parametric curve $(\alpha(q),f(q))$ or $(\alpha,f(\alpha))$ constitutes the multifractal spectrum and $\Delta \alpha$ (i.e., $\alpha_{max}-\alpha_{min}$) constitutes the multifractal spectrum width. $r$ determines only an adequately strong part of the multifractal spectrum is considered. This tradition of using a correlation-coefficient benchmark began with Dixon and Kelty-Stephen \cite{dixon2012multifractal} trying to operationalize the concerns raised by Zamir \cite{zamir2003critique}, and all Kelty-Stephen-co-authored empirical work using Chhabra and Jensen’s \cite{chhabra1989direct} multifractal analysis since has used this same benchmark. The use of correlation coefficient has regularly provided multifractal spectra whose widths have been significant predictors of various behavioral outcomes \cite{bell2019non,bloomfield2021perceiving,carver2017multifractal,carver2017multifractality,jacobson2021multifractality,kelty2016multifractal,kelty2021multifractality,mangalam2020multifractal,mangalam2020multiplicative,stephen2011strong,stephen2012multifractal,teng2016non,ward2018bringing}. So, whatever may be arbitrary in this choice of correlation coefficient and whatever may be determined alternatively/equally/more correctly or usefully, the systematic application of this standard across an entire series of experimental datasets has not left the estimated measures of multifractal spectrum width altogether useless.

\subsubsection{Assessing multifractality due to nonlinearity using surrogate testing}

To identify whether a nonzero $\Delta\alpha$ reflects multifractality due to cascade-like interactivity, $\Delta\alpha$ for the original series was compared to $\Delta\alpha$ for 32 Iterated Amplitude Adjusted Fourier Transform (IAAFT) surrogates \cite{ihlen2012introduction,schreiber1996improved}. IAAFT randomizes original values time-symmetrically around the autoregressive structure, generating surrogates that randomize phase ordering of the original series’ spectral amplitudes while preserving linear temporal correlations. The one-sample \textit{t}-statistic (henceforth, $t_{MF}$) takes the subtractive difference between $\Delta\alpha$ for the original series and the 32 surrogates, dividing by the standard error of the spectrum width for the 32 surrogates. The greater the value of $t_{MF}$, the greater the amount of multifractality in the original series due to nonlinear as opposed to linear sources.

\subsection{Estimating ergodicity breaking parameter \boldmath$\mathrm{EB}$ for FBM, SBM, nCTRW, ATTM, and LW series, and the corresponding \textbf{TA-}\boldmath$\mathrm{MSD}$, \boldmath$SD$, \boldmath$CV$, \boldmath$RMS$, \boldmath$H_{fGn}$, \boldmath$\Delta \alpha$, and \boldmath$t_{MF}$ series}

Ergodicity can be quantified using a dimensionless statistic of ergodicity breaking $\mathrm{EB}$, also known as the Thirumalai-Mountain metric \cite{he2008random,thirumalai1989ergodic} and already mentioned by Rytov et al. \cite{rytov1989principles}, computed by subtracting the squared total-sample variance from the average squared subsample variance and dividing the resultant by the squared total-sample variance:
\begin{equation*}
  \mathrm{EB}(x(t)) = \frac{ \Bigl \langle \Bigl [ \overline{\delta^{2}(x(t))} \Bigl ]^{2} \Bigl \rangle - \Bigl \langle \overline{\delta^{2}(x(t))} \Bigl \rangle^{2}}{ \Bigl \langle \overline{\delta^{2}(x(t))} \Bigl \rangle ^{2}}. \tag{16}\label{eq:16}
\end{equation*}
Rapid decay of $\mathrm{EB}$ to $0$ for progressively larger samples, i.e., $\mathrm{EB}\rightarrow0$ as $t \rightarrow \infty$ implies ergodicity. Thus, for Brownian motion $\mathrm{EB} (x(t)) = \frac{4}{3} (\frac{\Delta}{t})$ \cite{cherstvy2013anomalous,metzler2014anomalous}. Slower decay indicates less ergodic systems in which trajectories are less reproducible, and no decay or convergence to a finite asymptotic value indicates strong ergodicity breaking \cite{deng2009ergodic}. $\mathrm{EB}$-vs.-$t$ curves thus allow testing whether a given time series fulfills ergodic assumptions or breaks ergodicity and the extent to which it breaks ergodicity.

For instance, Deng and Barkai \cite{deng2009ergodic} have shown that for FBM,
\begin{equation*}
  \mathrm{EB}(x(t)) =
  \begin{cases}
    k (H_{fGn}) \frac{\Delta}{t} & \text{if } 0 < H_{fGn} < \frac{3}{4} \\
    k (H_{fGn}) \frac{\Delta}{t}\ln{t} & \text{if } H_{fGn} = \frac{3}{4} \\
    k (H_{fGn}) ( \frac{\Delta}{t} ) ^{4 - 4H_{fGn}} & \text{if } \frac{3}{4} < H_{fGn} < 1.
  \end{cases}
  \tag{17}\label{eq:17}
\end{equation*}
Likewise, Thiel and Sokolov \cite{thiel2014scaled} have shown that for SBM,
\begin{equation*}
\mathrm{EB}(x(t)) =
  \begin{cases}
  4Z_{\alpha} \biggl( \frac{\Delta}{t} \biggl) ^{2 \alpha} & \text{if } \alpha \leq \frac{1}{2} \\
  \frac{4\alpha^2}{3(2\alpha - 1)}\frac{\Delta}{t} & \text{if } \alpha > \frac{1}{2},
  \end{cases}
  \tag{18}\label{eq:18}
\end{equation*}
where $Z_{\alpha}= \int_{0}^{1} dy \int_{0}^{\infty} dx[(x+1)^{\alpha}-(x+y)^{\alpha}]^2$. For any positive value of the anomalous exponent $\alpha$, $\mathrm{EB}$ for SBM vanishes and shows a crossover between two types of $t$-dependence at $\alpha = 1/2$ \cite{thiel2014scaled}. Compare also \cite{rebenshtok2007distribution}. Despite the vanishing $\mathrm{EB}$, SBM is weakly non-ergodic, i.e., ensemble and time averages are disparate.

The present work is less focused on firmly meeting the criterion of $\mathrm{EB}$ converging to zero or a very low finite value within our finite samples. Instead, we aim to compare the decay rate in $\mathrm{EB}$ in FBM, SBM, nCTRW, ATTM, and LW, and the corresponding $\varA{TA-MSD}$, $SD$, $CV$, $RMS$, $H_{fGn}$, $\Delta \alpha$, and $t_{MF}$ series. We used the comparison among anomalous diffusion processes as a measure of ergodicity breaking instead of strict convergence to zero. We computed $\mathrm{EB}$ for the unsigned series obtained for the original and a shuffled version of each process (range = $T/50$; lag $\Delta$ = 10 samples), for $\varA{TA-MSD}$ for the original and a shuffled version of each process (range = $T/50$; lag $\Delta$ = 10 samples), and for each $\varA{TA-MSD}$, $SD$, $CV$, $RMS$, $H_{fGn}$, $\Delta\alpha$, and $t_{MF}$ series computed over the 100 non-overlapping segments for the original and a shuffled version of each process (range = $s/2$; lag $\Delta$ = 1 segment).

\section{Results}

\subsection{Ergodicity breaking depends on the type of the diffusion process and the anomalous exponent \boldmath$\alpha$}

We observe noteworthy ergodicity-related differences across the five types of anomalous diffusion processes and different values of the anomalous exponent $\alpha$. FBM for smaller values of $\alpha$ (i.e., $\alpha \rightarrow 0.1$) and SBM for larger values of $\alpha$ (i.e., $\alpha \rightarrow 1$) return and converge towards the mean over a bigger ensemble, suggestive of ergodicity (Fig.~\ref{fig:f2}a). The respective $\mathrm{EB}$-vs.-$t$ curves confirmed this observation. For FBM, $\mathrm{EB} \rightarrow 0$ as $\alpha \rightarrow 0.1$, and for SBM, $\mathrm{EB} \rightarrow 0$ as $\alpha \rightarrow 1$ (Fig.~\ref{fig:f2}b). FMB appears to break ergodicity for larger values of $\alpha$, indicated by little to no decay in $\mathrm{EB}$ with $t$. SBM appears to break ergodicity for smaller values of $\alpha$, which, despite decay in $\mathrm{EB}$ with $t$, do not even reach $1$. Ergodicity in FBM and SBM is further confirmed by the observation that the $\mathrm{EB}$-vs.-$t$ curves almost entirely coincide for the original and shuffled trajectories as $\alpha \rightarrow 0.1$ for FBM and $\alpha \rightarrow 1$ for SBM, indicating that these processes behaved as awGn. In contrast, the $\mathrm{EB}$-vs.-$t$ curves for the original and shuffled versions coincide to progressively lesser extent as $\alpha \rightarrow 1$ for FMB and $\alpha \rightarrow 0.1$ for SBM. Overall, not only $\mathrm{EB}$ for the five processes does not converge to an acceptably small value within our observation time, the FBM, SBM, nCTRW, ATTM, and LW trajectories show highly variable rates of decays in $\mathrm{EB}$: $\mathrm{EB}(x(t)) = -1.01\frac{\Delta}{s}\;\mathrm{to}-0.18\frac{\Delta}{s}, -1.05\frac{\Delta}{s}\;\mathrm{to}-0.42\frac{\Delta}{s}, -1.11\frac{\Delta}{s}\;\mathrm{to}-0.25\frac{\Delta}{s}, -1.87\frac{\Delta}{s}\;\mathrm{to}-0.45\frac{\Delta}{s},\mathrm{and}-1.02\frac{\Delta}{s}\;\mathrm{to}-0.99\frac{\Delta}{s}$, respectively, where $\Delta\;\mathrm{in}\frac{\Delta}{t} = 10$, a not atypical value for many empirical data.

Unlike FBM and SBM, nCTRW, ATTW, and LW diverge and never return towards the mean over a bigger ensemble, suggestive of ergodicity breaking in these processes (Fig.~\ref{fig:f2}a). $\mathrm{EB}$ quickly decays with $t$ but does not even reach $1$ for nCTRW and ATTW for smaller values of $\alpha$ (i.e., $\alpha \rightarrow 0.1$) and for LW for all values of $\alpha$, suggestive of weak ergodicity breaking (Fig.~\ref{fig:f2}b). $\mathrm{EB}$ showed little to no decay with $t$ for nCTRW and ATTW for larger values of $\alpha$ (i.e., $\alpha \rightarrow 1$), suggestive of stronger breaking of ergodicity. This ergodicity breaking of nCTRW and ATTW is further confirmed by the finding that the $\mathrm{EB}$-vs.-$t$ curves for the original and shuffled nCTRW and ATTW coincide to progressively lesser extent as $\alpha \rightarrow 1$. LW showed the weakest ergodicity breaking, wherein $\mathrm{EB}$ quickly decays with $t$ for all values of $\alpha$ but does not reach $1$. Hence, FBM and SBM apparently break ergodicity for larger and smaller values of the anomalous exponent $\alpha$, nCTRW, ATTM, and SBM break ergodicity to differential extents.

\subsection{\boldmath$\varA{TA-MSD}$ reflects ergodicity-related differences among different anomalous diffusion processes}

\begin{figure*}
\includegraphics{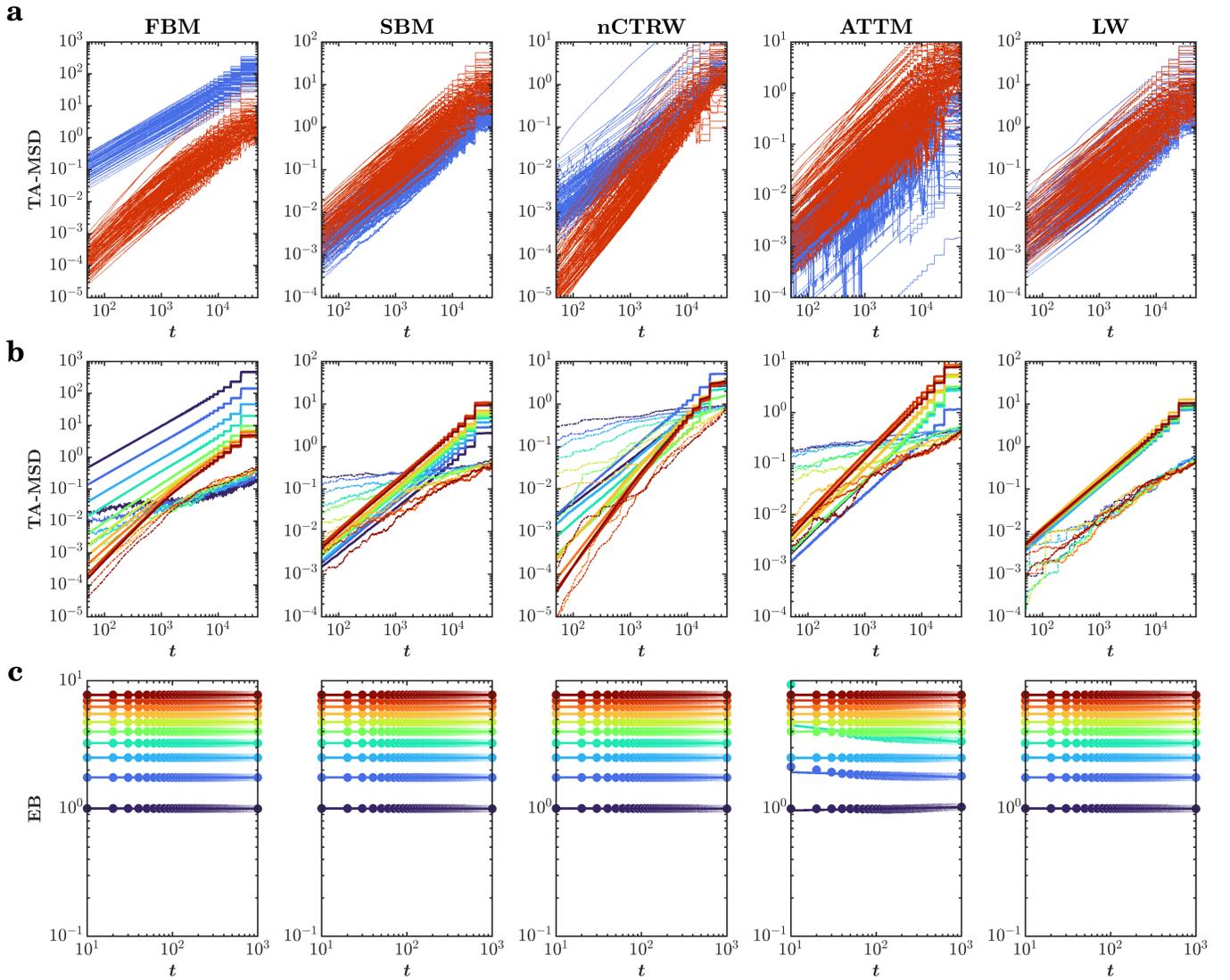}
\caption{Ergodicity breaking in $\varA{TA-MSD}$. (a) $\varA{TA-MSD}$ for 100 simulated trajectories of the five anomalous diffusion processes—FBM, SBM, nCTRW, ATTM, and LW, for two values of the anomalous exponent $\alpha$ (blue: $\alpha$ = 0.2 for nCTRW, ATTM, FBM, and SBM, and $\alpha$ = 1.2 for LW; red: $\alpha$ = 0.9 for nCTRW, ATTM, FBM, and SBM, and $\alpha$ = 1.9 for LW. (b) $\varA{EA-MSD}$ (colored solid lines) and $\varA{TA-MSD}$ (colored dash-dotted lines) for the five anomalous diffusion processes—FBM, SBM, nCTRW, ATTM, and LW. The anomalous exponent $\alpha$ ranges from 0.1 to 1 for FBM, SBM, nCTRW, and ATTM, and from 1.1 to 2 for LW, with an increment of 0.1 from dark blue to dark red. (c) $\mathrm{EB}$-vs.-$s$ curves for $\varA{TA-MSD}$ for the five processes and different values of $\alpha$ ($N = 100$; lag is $\Delta = 10$ samples). The $\mathrm{EB}$-vs.-$s$ curves all coincide at $10^{0}$ but have been shifted vertically for convenience of presentation, and hence, the vertical axis is given in arbitrary units.}
\label{fig:f3}
\end{figure*}

Figs.~\ref{fig:f3}a \& \ref{fig:f3}b show that $\varA{TA-MSD}$ grows with lag time $\Delta$ for all five types of anomalous diffusion processes—FBM, SBM, nCTRW, ATTM, and LW—in the entire range of $\Delta$. (The individual amplitudes scatter owing to the stochasticity in the generation of these trajectories. Such scatter characteristics of anomalous diﬀusion processes can be used to reliably distinguish between FBM from nCTRW processes, for instance, \cite{burov2011single,jeon2010analysis}.) Furthermore, the growth of $\varA{TA-MSD}$ shows differential dependence on the anomalous exponent $\alpha$ across the five processes. For example, this dependence was strongest for FBM and lowest for LW, suggesting that the growth rates of $\varA{TA-MSD}$ might not provide sufficient resolution to distinguish multiple trajectories of the same process with different anomalous exponents. $\mathrm{EB}$ for $\varA{TA-MSD}$ remains constant across $t$ (Fig.~\ref{fig:f3}c); indeed, $\varA{TA-MSD}$ for FBM, SBM, nCTRW, ATTM, and LW show no decay whatsoever in $\mathrm{EB}$: $\mathrm{EB}(\varA{TA-MSD}(t)) = 0$ for all values of the anomalous exponent $\alpha$. Hence, $\varA{TA-MSD}$ mask any ergodicity-related differences among the five processes and processes with different anomalous exponents and cannot be used as a stable causal predictor in the linear modeling of cause-effect relationships in this type of analysis.

\subsection{\boldmath$\varA{MSD}$-related linear descriptors such as \boldmath$SD$, \boldmath$CV$, and \boldmath$RMS$ reflect ergodicity-related differences among different anomalous diffusion processes}

\begin{figure*}
\includegraphics{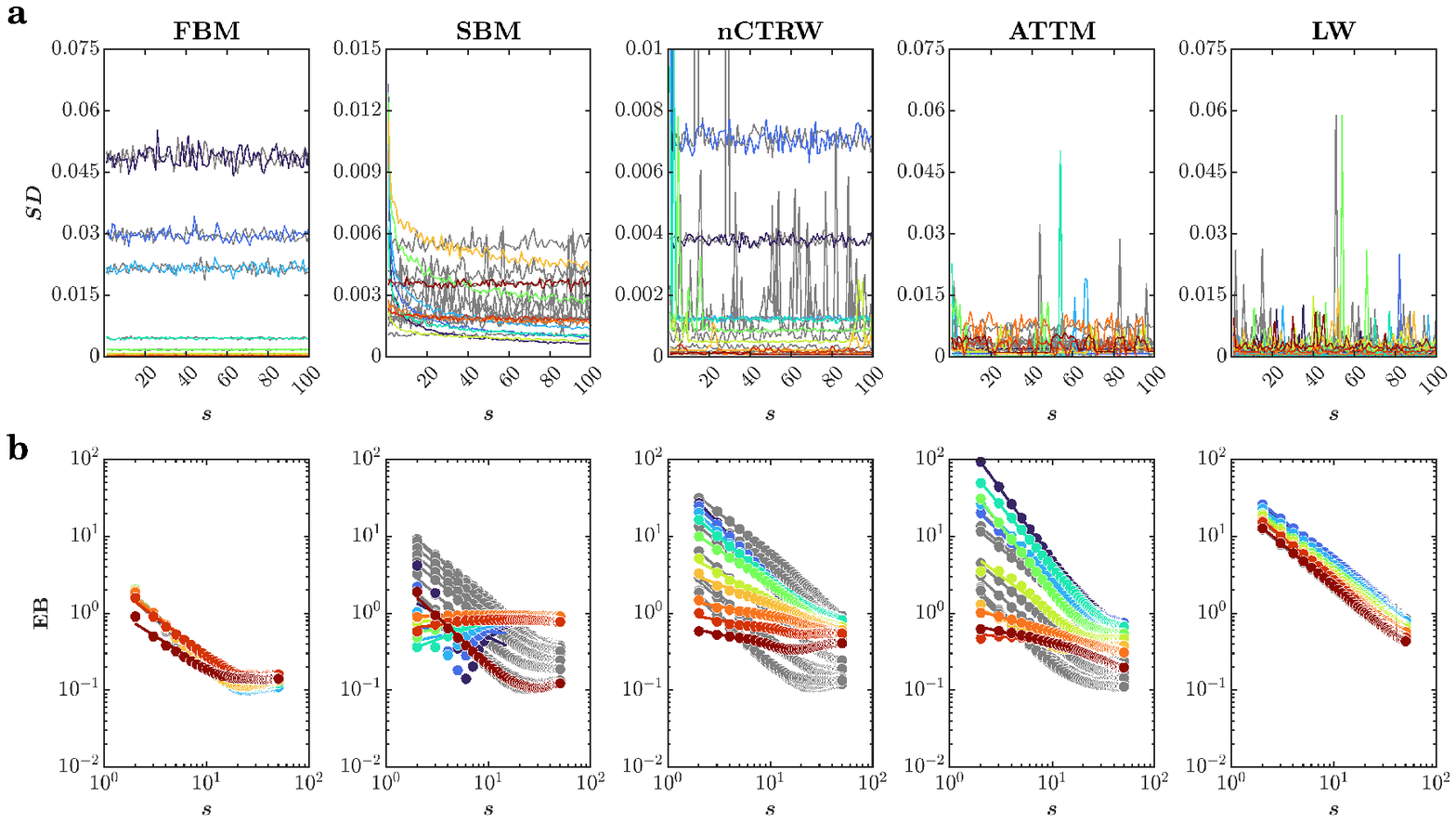}
\caption{Ergodicity breaking in $SD$. (a) Representative $SD$ series ($SD$ calculated across the 100 non-overlapping 500-sample segments) for the five anomalous diffusion processes—FBM, SBM, nCTRW, ATTM, and LW. The anomalous exponent $\alpha$ ranges from 0.1 to 1 for FBM, SBM, nCTRW, and ATTM, and from 1.1 to 2 for LW, with an increment of 0.1 from dark blue to dark red. Grey trajectories indicate $SD$ series for the corresponding shuffled versions. (b) $\mathrm{EB}$-vs.-$s$ curves for $SD$ series for the five processes and different values of $\alpha$ ($N = 100$; lag is $\Delta = 1$ segment). Grey curves indicate mean $\mathrm{EB}$-vs.-$s$ curves for the corresponding shuffled versions.}
\label{fig:f4}
\end{figure*}

$SD$ for FBM for all values of the anomalous exponent $\alpha$ and SBM for $\alpha = 1$—the particular case in which SBM is reduced to awGn—return and converge towards the mean over a bigger ensemble, suggestive of ergodicity (Fig.~\ref{fig:f4}a). In contrast, $SD$ for SBM, nCTRW, ATTW, and LW diverge and never return towards the mean over a bigger ensemble, suggesting ergodicity breaking in the $SD$ series for these processes. The ergodicity breaking parameter vs. segment curves, i.e., $\mathrm{EB}$-vs.-$s$ curves, confirmed these trends. $\mathrm{EB} \rightarrow 0$ as $s \rightarrow \infty$ for FBM for all values of $\alpha$ and SBM for $\alpha = 1$—the particular case in which SBM is reduced to awGn (Fig.~\ref{fig:f4}b). $\mathrm{EB}$ shows no decay with $s$ for SBM, a much slower decay with $s$ for nCTRW, and a quick decay but to a much larger value for ATTM and LW, especially for larger values of $\alpha$ (i.e, $\alpha \rightarrow 1$), confirming ergodicity breaking in $SD$ for these processes. Overall, the $SD$ series for FBM, SBM, nCTRW, ATTM, and LW show highly variable initial rates of decay in $\mathrm{EB}$: $\mathrm{EB}(SD(s)) = -1.32\frac{\Delta}{s}\;\mathrm{to}-0.84\frac{\Delta}{s}, -1.25\frac{\Delta}{s}\;\mathrm{to}0.48\frac{\Delta}{s}, -1.49\frac{\Delta}{s}\;\mathrm{to}-0.23\frac{\Delta}{s}, -1.83\frac{\Delta}{s}\;\mathrm{to}-1.16\frac{\Delta}{s},\mathrm{and}-1.11\frac{\Delta}{s}\;\mathrm{to}-1.04\frac{\Delta}{s}$, respectively, where $\Delta = 1$. In other words, the $SD$ series for the five processes show highly variable rates of decay in $\mathrm{EB}$.

\begin{figure*}
\includegraphics{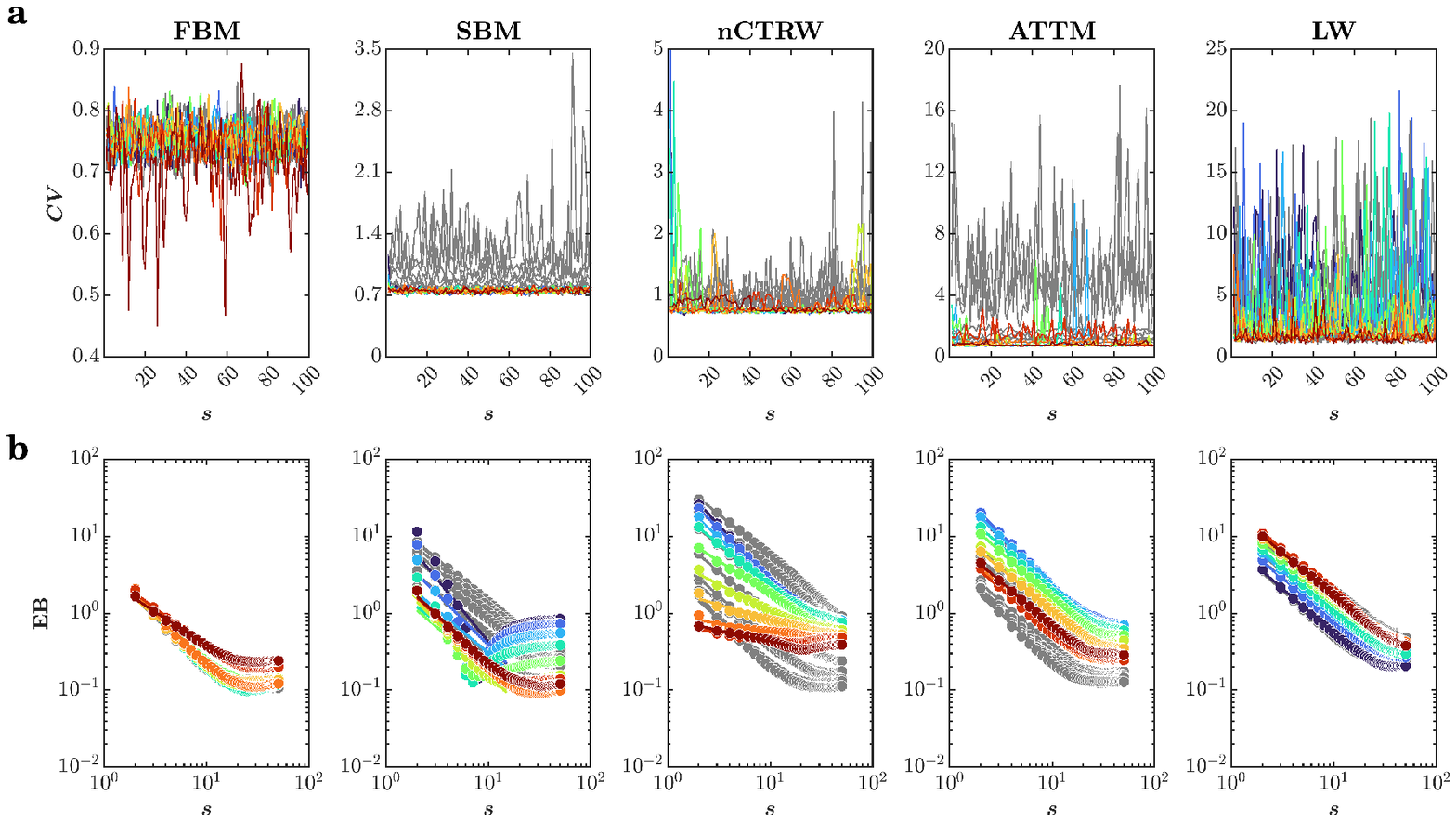}
\caption{Ergodicity breaking in $CV$. (a) Representative $CV$ series ($CV$ calculated across the 100 non-overlapping 500-sample segments) for the five anomalous diffusion processes—FBM, SBM, nCTRW, ATTM, and LW. The anomalous exponent $\alpha$ ranges from 0.1 to 1 for FBM, SBM, nCTRW, and ATTM, and from 1.1 to 2 for LW, with an increment of 0.1 from dark blue to dark red. Grey trajectories indicate $CV$ series for the corresponding shuffled versions. (b) $\mathrm{EB}$-vs.-$s$ curves for $CV$ series for the five processes and different values of $\alpha$ ($N = 100$; lag is $\Delta = 1$ segment). Grey curves indicate mean $\mathrm{EB}$-vs.-$s$ curves for the corresponding shuffled versions.}
\label{fig:f5}
\end{figure*}

Except for some minor differences, the ergodicity breaking behavior of $CV$ resembles that of $SD$. $CV$ for FBM return and converge towards the mean for all values of the anomalous exponent $\alpha$ and SBM for $\alpha = 1$—the particular case in which SBM is reduced to awGn—return and converge towards the mean over a bigger ensemble, suggestive of ergodicity (Fig.~\ref{fig:f5}a). In contrast, $CV$ for SBM, nCTRW, ATTW, and LW diverge and never return towards the mean over a bigger ensemble, suggesting ergodicity breaking in the $CV$ series for these processes. The $\mathrm{EB}$-vs.-$s$ curves confirmed these trends. $\mathrm{EB} \rightarrow 0$ as $s \rightarrow \infty$ for FBM for the most part but show marginal ergodicity breaking for larger values of $\alpha$ (Fig.~\ref{fig:f5}b). $CV$ series for SBM behaved ergodically for $\alpha = 1$—the particular case in which SBM is reduced to awGn, but $CV$ series for SBM for other values of $\alpha$ showed strong ergodicity breaking. $CV$ series for nCTRW break ergodicity, with stronger ergodicity breaking for larger $\alpha$. $CV$ series for ATTM and LW also show marginal ergodicity breaking, wherein $\mathrm{EB}$ quickly decayed with $s$ but did not reach 1. Overall, the $CV$ series for FBM, SBM, nCTRW, ATTM, and LW show highly variable initial rates of decay in $\mathrm{EB}$: $\mathrm{EB}(CV(s)) = -1.33\frac{\Delta}{s}\;\mathrm{to}-0.87\frac{\Delta}{s}, -1.62\frac{\Delta}{s}\;\mathrm{to}-1.02\frac{\Delta}{s}, -1.49\frac{\Delta}{s}\;\mathrm{to}-0.20\frac{\Delta}{s}, -1.51\frac{\Delta}{s}\;\mathrm{to}-1.09\frac{\Delta}{s},\mathrm{and}-1.15\frac{\Delta}{s}\;\mathrm{to}-1.06\frac{\Delta}{s}$, respectively, where $\Delta = 1$. In other words, the $CV$ series for the five processes show highly variable rates of decay in $\mathrm{EB}$.

\begin{figure*}
\includegraphics{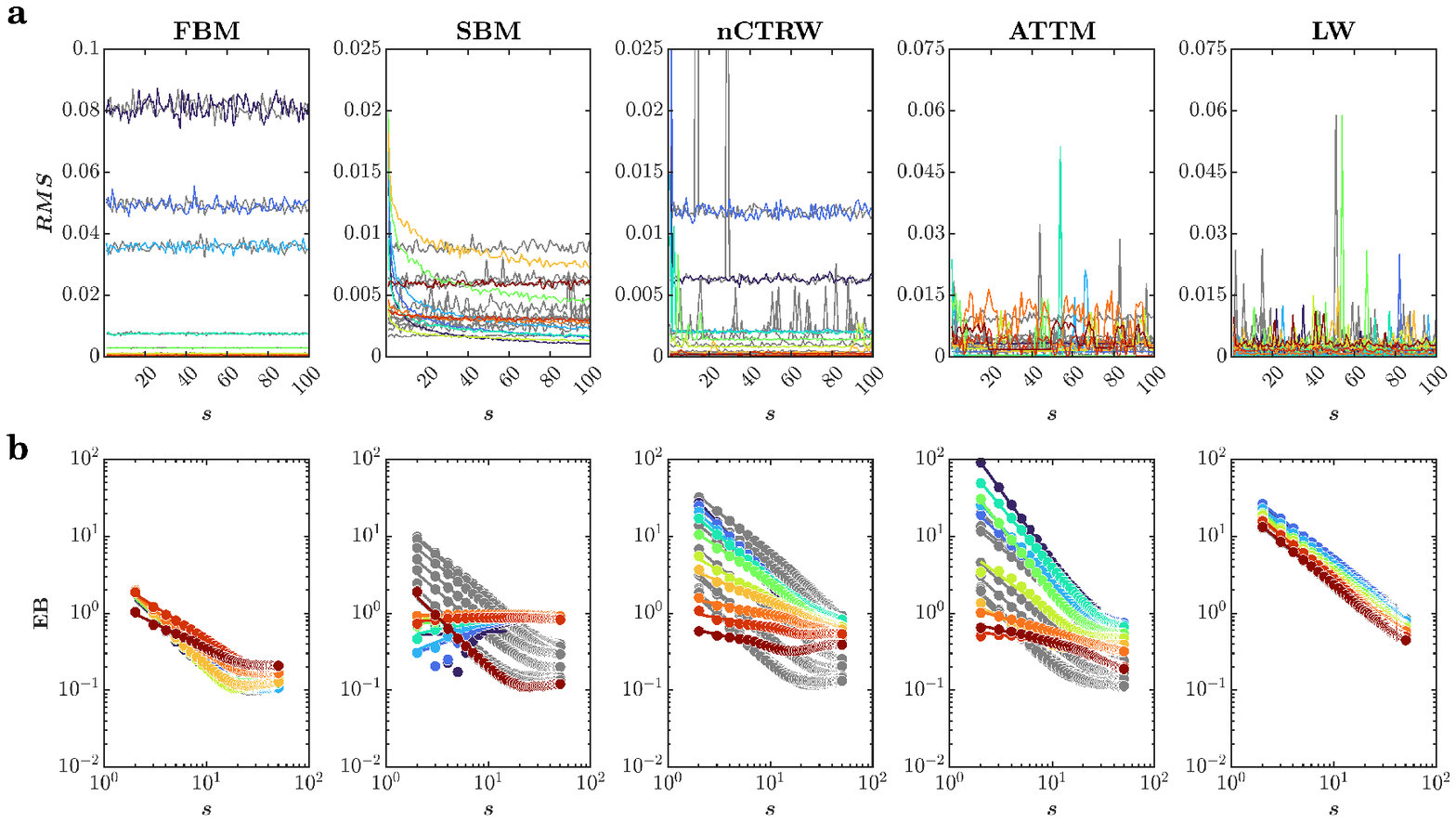}
\caption{Ergodicity breaking in $RMS$. (a) Representative $RMS$ series ($RMS$ calculated across the 100 non-overlapping 500-sample segments) for the five anomalous diffusion processes—FBM, SBM, nCTRW, ATTM, and LW. The anomalous exponent $\alpha$ ranges from 0.1 to 1 for FBM, SBM, nCTRW, and ATTM, and from 1.1 to 2 for LW, with an increment of 0.1 from dark blue to dark red. Grey trajectories indicate $RMS$ series for the corresponding shuffled versions. (b) $\mathrm{EB}$-vs.-$s$ curves for $RMS$ series for the five processes and different values of $\alpha$ ($N = 100$; lag is $\Delta = 1$ segment). Grey curves indicate mean $\mathrm{EB}$-vs.-$s$ curves for the corresponding shuffled versions.}
\label{fig:f6}
\end{figure*}

$RMS$ series behaved exactly as $SD$ series for all five processes. $RMS$ for FBM for all values of the anomalous exponent $\alpha$ and SBM for $\alpha = 1$—the particular case in which SBM is reduced to awGn—return and converge towards the mean over a bigger ensemble, suggestive of ergodicity (Fig.~\ref{fig:f6}a). In contrast, $RMS$ for SBM, nCTRW, ATTW, and LW diverge and never return towards the mean over a bigger ensemble, suggesting ergodicity breaking in the $RMS$ series for these processes. The $\mathrm{EB}$-vs.-$s$ curves confirmed these trends. $\mathrm{EB} \rightarrow 0$ as $s \rightarrow \infty$ for FBM for all values of $\alpha$ and SBM for $\alpha = 1$—the particular case in which SBM is reduced to awGn (Fig.~\ref{fig:f6}b). $\mathrm{EB}$ shows no decay with $s$ for SBM, a much slower decay with $s$ for nCTRW, and a quick decay but to a much larger value for ATTM and LW, especially for larger values of $\alpha$ (i.e, $\alpha \rightarrow 1$), confirming ergodicity breaking in $RMS$ for these processes. Overall, the $RMS$ series for FBM, SBM, nCTRW, ATTM, and LW show highly variable initial rates of decay in $\mathrm{EB}$: $\mathrm{EB}(RMS(s)) = -1.36\frac{\Delta}{s}\;\mathrm{to}0.64\frac{\Delta}{s}, -1.23\frac{\Delta}{s}\;\mathrm{to}-1.00\frac{\Delta}{s}, -1.28\frac{\Delta}{s}\;\mathrm{to}-0.53\frac{\Delta}{s}, -1.84\frac{\Delta}{s}\;\mathrm{to}-0.18\frac{\Delta}{s},\mathrm{and}-1.12\frac{\Delta}{s}\;\mathrm{to}-1.04\frac{\Delta}{s}$, respectively, where $\Delta = 1$. In other words, the $RMS$ series for the five processes show highly variable rates of decay in $\mathrm{EB}$.

\subsection{Multifractal descriptors provide an ergodic characterization of non-ergodic anomalous diffusion processes}

\begin{figure*}
\includegraphics{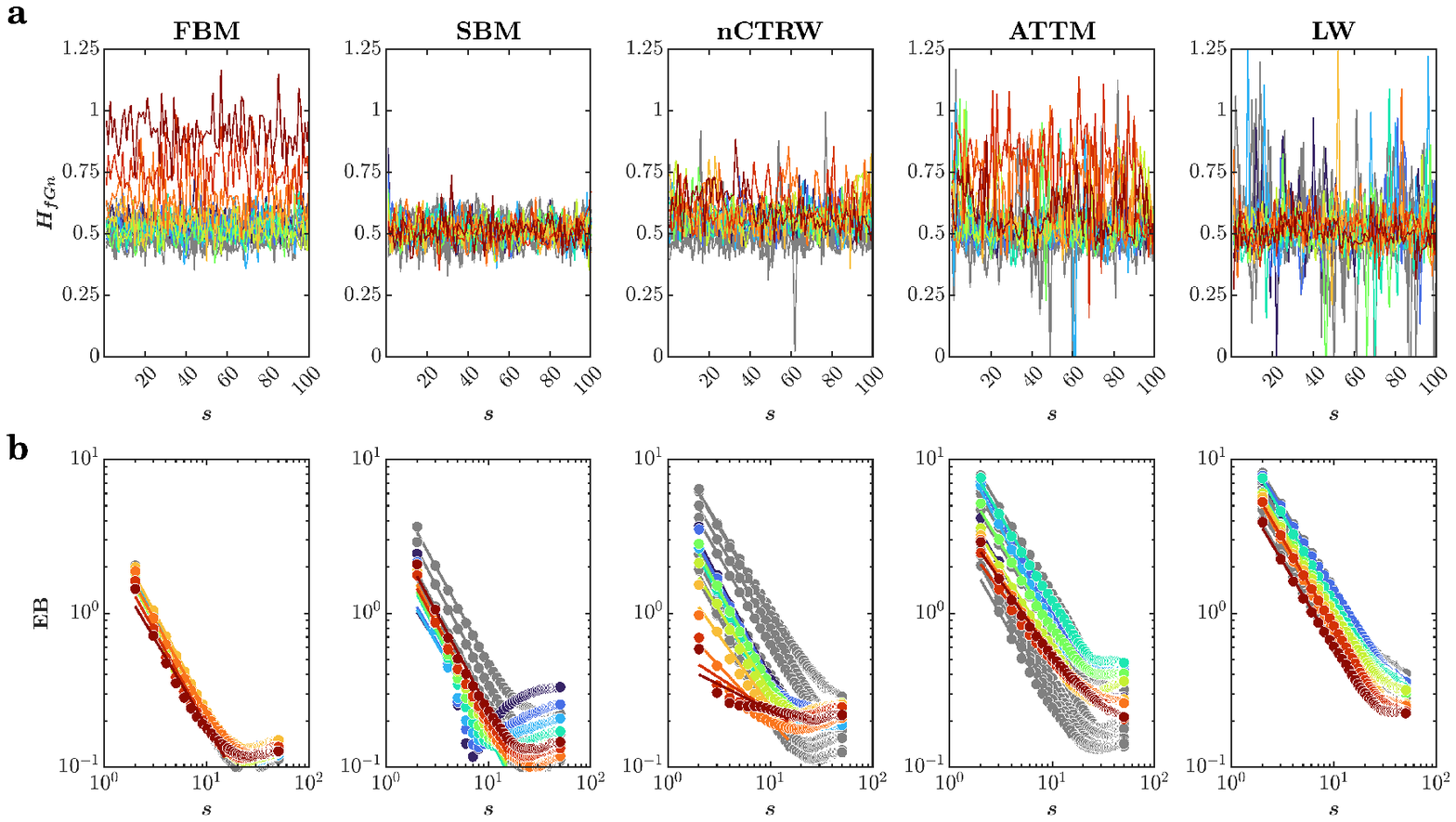}
\caption{Ergodicity breaking in the strength of temporal correlations, $H_{fGn}$. (a) Representative $H_{fGn}$ series ($H_{fGn}$ calculated across the 100 non-overlapping 500-sample segments) for the five anomalous diffusion processes—FBM, SBM, nCTRW, ATTM, and LW. The anomalous exponent $\alpha$ ranges from 0.1 to 1 for FBM, SBM, nCTRW, and ATTM, and from 1.1 to 2 for LW, with an increment of 0.1 from dark blue to dark red. Grey trajectories indicate $H_{fGn}$ series for the corresponding shuffled versions. (b) $\mathrm{EB}$-vs.-$s$ curves for $H_{fGn}$ series for the five processes and different values of $\alpha$ ($N = 100$; lag is $\Delta = 1$ segment). Grey curves indicate mean $\mathrm{EB}$-vs.-$s$ curves for the corresponding shuffled versions.}
\label{fig:f7}
\end{figure*}

The $H_{fGn}$ series---quantifying the strength of temporal correlations in each of the 100 non-overlapping 500-sample segments of the synthetic trajectories---shows signs of restoring ergodicity to the description of some of these processes. The $H_{fGn}$ series for FBM behaved ergodically (i.e., $\mathrm{EB} \rightarrow 0$ as $s \rightarrow \infty$) independent of the anomalous exponent $\alpha$, and so do $H_{fGn}$ series for LW (Fig.~\ref{fig:f7}). However, the $H_{fGn}$ series for LW show a marginal dependence on $\alpha$, as the $\mathrm{EB}$ taper off at higher values of $s$. This result is further strengthened by the observation that these $\mathrm{EB}$-vs.-$s$ curves entirely coincide with the original and shuffled trajectories. The $H_{fGn}$ series for SBM show strong ergodicity breaking with $\mathrm{EB}$ initially having a converging-to-zero trend but then taking an upward turn and increasing with $s$ for the rest of the range. The only exception to this trend is the $H_{fGn}$ series for the SBM trajectories for $\alpha = 1$—the particular case in which SBM is reduced to awGn. The $H_{fGn}$ series for CTRW and ATTM also behaved ergodically with a few exceptions: the $H_{fGn}$ series for CTRW break ergodicity at larger values of $\alpha$ and the $H_{fGn}$ series for the original and shuffled ATTM trajectories diverge with increasing $\alpha$. On average, the $H_{fGn}$ series for FBM, SBM, nCTRW, ATTM, and LW show the initial rates of decay in $\mathrm{EB}$: $\mathrm{EB}(H_{fGn}(s)) = -1.23\frac{\Delta}{s}, -1.23\frac{\Delta}{s}, -0.97\frac{\Delta}{s}, -1.08\frac{\Delta}{s},\mathrm{and}-1.12\frac{\Delta}{s}$, respectively, where $\Delta = 1$. In other words, the $H_{fGn}$ series for the five processes show very similar rates of decay in $\mathrm{EB}$. 

\begin{figure*}
\includegraphics{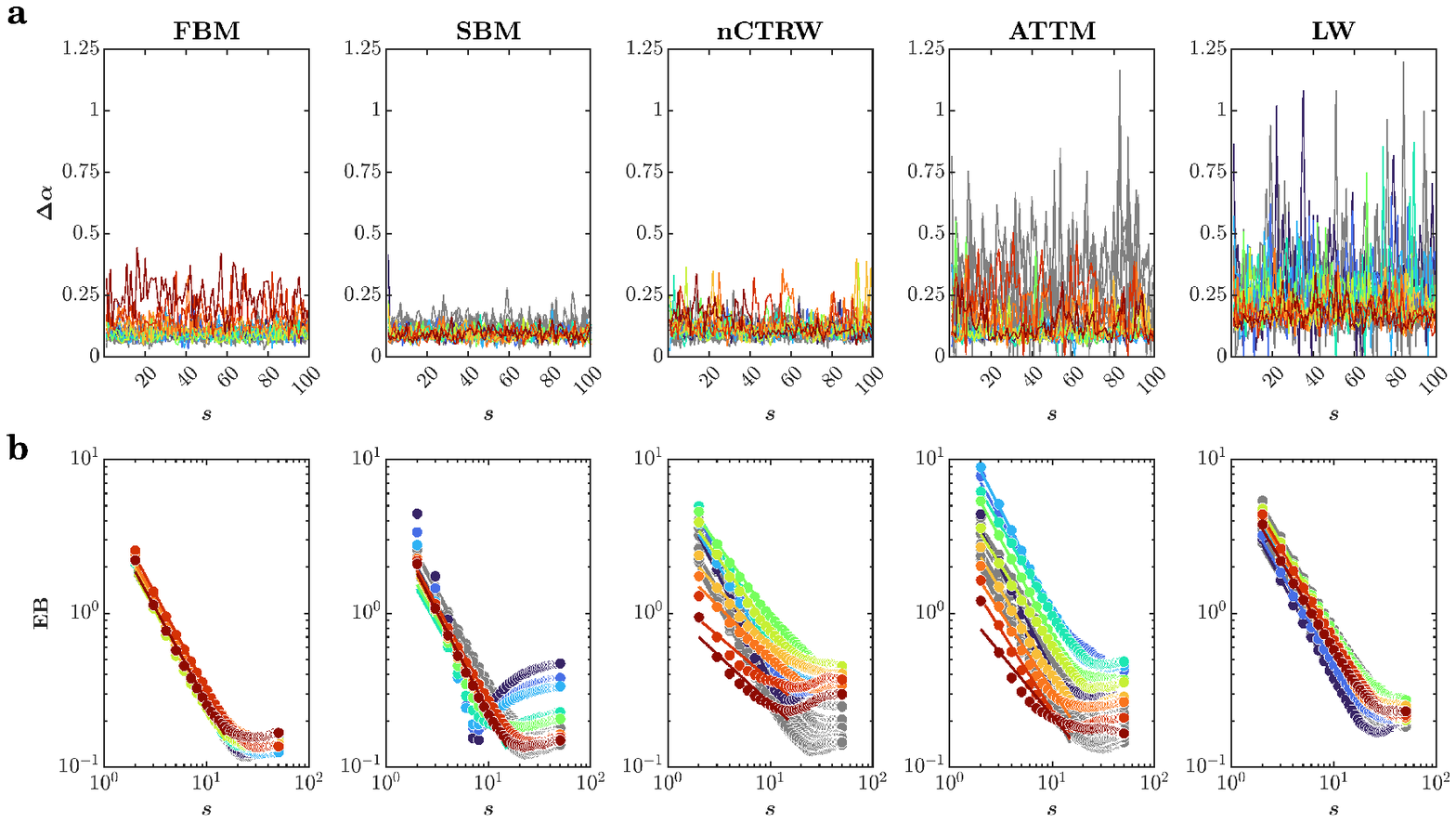}
\caption{Ergodicity breaking in multifractal spectrum width, $\Delta \alpha$. (a) Representative $\Delta \alpha$ series ($\Delta \alpha$ calculated across the 100 non-overlapping 500-sample segments) for the five anomalous diffusion processes—FBM, SBM, nCTRW, ATTM, and LW. The anomalous exponent $\alpha$ ranges from 0.1 to 1 for FBM, SBM, nCTRW, and ATTM, and from 1.1 to 2 for LW, with an increment of 0.1 from dark blue to dark red. Grey trajectories indicate $\Delta \alpha$ series for the corresponding shuffled versions. (b) Mean $\mathrm{EB}$-vs.-$s$ curves for $\Delta \alpha$ series for the five processes and different values of $\alpha$ ($N = 100$; lag is $\Delta = 1$ segment). Grey curves indicate mean $\mathrm{EB}$-vs.-$s$ curves for the corresponding shuffled versions.}
\label{fig:f8}
\end{figure*}

The $\Delta \alpha$ series---quantifying the width of the multifractal spectrum in each of the 100 non-overlapping 500-sample segments of the synthetic trajectories---also show signs of restoring ergodicity to all five processes, albeit with some exceptions pertaining to specific values of the anomalous exponent. The $\Delta \alpha$ series for FBM behave ergodically (i.e., $\mathrm{EB} \rightarrow 0$ as $s \rightarrow \infty$) independent of $\alpha$, and so do the $\Delta \alpha$ series for LW (Fig.~\ref{fig:f8}). However, the $\Delta \alpha$ series for LW show marginal dependence on $\alpha$, as $\mathrm{EB}$ taper off at higher values of $s$. The $\Delta \alpha$ series for SBM show strong ergodicity breaking with $\mathrm{EB}$ initially having a converging-to-zero trend but then taking an upward turn and increasing with $s$ for the rest of the range. The only exception to this trend is the $\Delta \alpha$ series for the SBM trajectories for $\alpha = 1$—the particular case in which SBM is reduced to awGn. The $\Delta \alpha$ series for CTRW and ATTM also behave ergodically with a few exceptions: the $\Delta \alpha$ series for CTRW break ergodicity at larger values of $\alpha$, and the $\Delta \alpha$ series for the original and shuffled ATTM trajectories diverge with increasing $\alpha$. On average, the $\Delta \alpha$ series for FBM, SBM, nCTRW, ATTM, and LW show the average initial rates of decay in $\mathrm{EB}$: $\mathrm{EB}(\Delta \alpha(s)) = -1.23\frac{\Delta}{s}, -1.22\frac{\Delta}{s}, -0.93\frac{\Delta}{s}, -1.12\frac{\Delta}{s},\mathrm{and}-1.16\frac{\Delta}{s}$, respectively, where $\Delta = 1$.

\begin{figure*}
\includegraphics{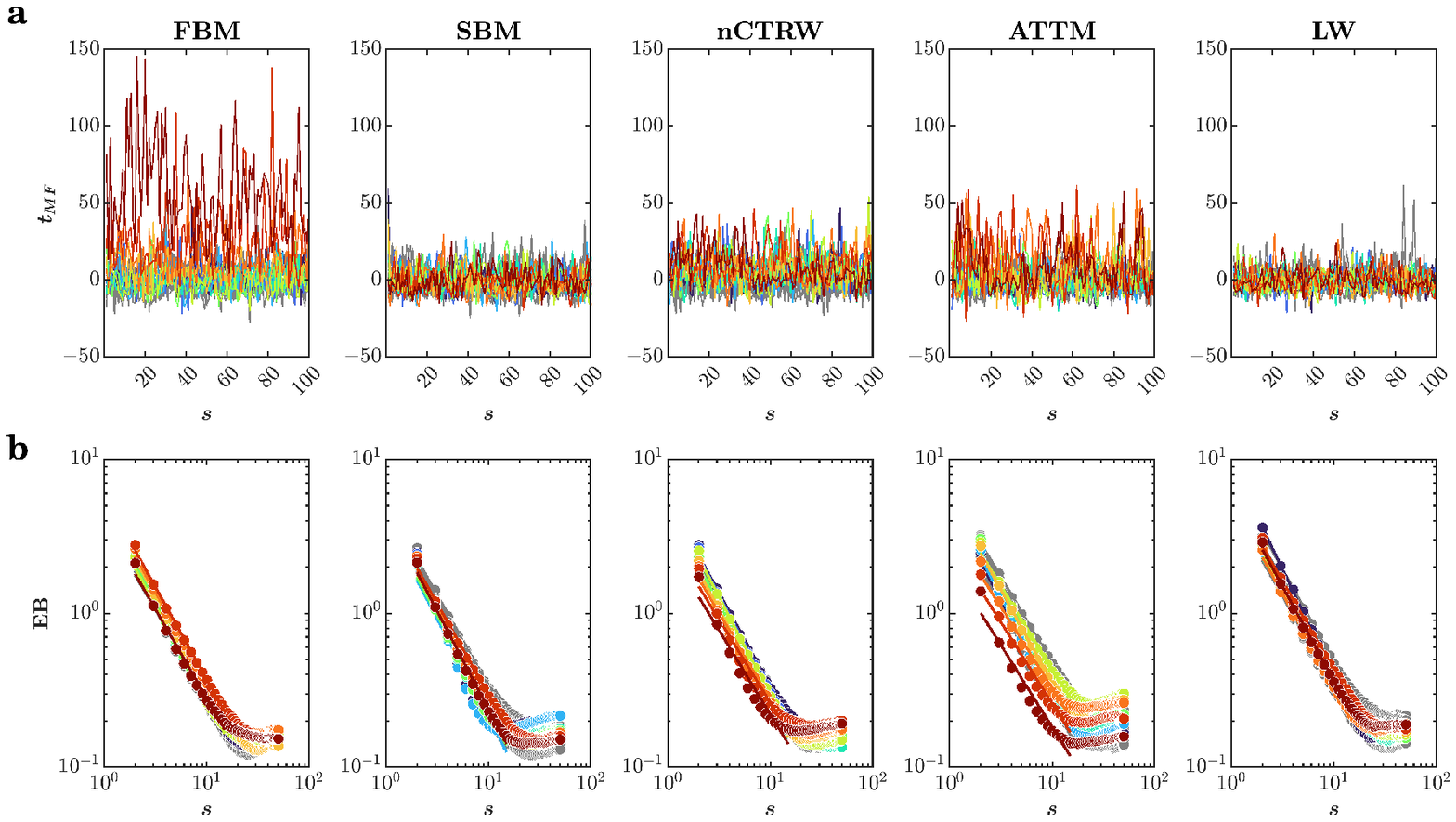}
\caption{Ergodicity breaking in multifractal nonlinearity, $t_{MF}$. (a) Representative $t_{MF}$ series ($t_{MF}$ calculated across the 100 non-overlapping 500-sample segments) for the five anomalous diffusion processes—FBM, SBM, nCTRW, ATTM, and LW. The anomalous exponent $\alpha$ ranges from 0.1 to 1 for FBM, SBM, nCTRW, and ATTM, and from 1.1 to 2 for LW, with an increment of 0.1 from dark blue to dark red. Grey trajectories indicate $t_{MF}$ series for the corresponding shuffled versions. (b) $\mathrm{EB}$-vs.-$s$ curves for $t_{MF}$ series for the five processes and different values of $\alpha$ ($N = 100$; lag is $\Delta = 1$ segment). Grey curves indicate mean $\mathrm{EB}$-vs.-$s$ curves for the corresponding shuffled versions.}
\label{fig:f9}
\end{figure*}

The $t_{MF}$ series---quantifying multifractality due to nonlinearity in each of the 100 non-overlapping 500-sample segments of the synthetic trajectories---for all five processes—FBM, SBM, nCTRW, ATTM, and LW—show a rapid decay of $\mathrm{EB}$ with a progressively larger sample of segments, i.e., $\mathrm{EB}\rightarrow0$ as $s \rightarrow0$ for $t\rightarrow\infty$ (Fig.~\ref{fig:f9}). On average, the $t_{MF}$ series for FBM, SBM, nCTRW, ATTM, and LW show the average initial rates of decay in $\mathrm{EB}$: $\mathrm{EB}(t_{MF}(s)) = -1.20\frac{\Delta}{s}, -1.24\frac{\Delta}{s}, -1.20\frac{\Delta}{s}, -1.17\frac{\Delta}{s},\mathrm{and}-1.20\frac{\Delta}{s}$, respectively, where $\Delta = 1$. In other words, the $t_{MF}$ series for the five processes do not vary in the initial rates of decay in $\mathrm{EB}$. Moreover, the $\mathrm{EB}$-vs.-$s$ curves show a marginal dependence on the anomalous exponent $\alpha$ at higher values of $s$, except for ATTM for which the $\mathrm{EB}$-vs.-$s$ curves show marginally higher dependence on $\alpha$. Notably, the $\mathrm{EB}$- vs.-$s$ curves entirely coincide for the original and shuffled trajectories, demonstrating that the ergodic behavior of the $t_{MF}$ series for the original series did not differ from the shuffled versions that lack any temporal correlations found in the original trajectories. Hence, the $t_{MF}$ series fully restore broken ergodicity to a description of all five diffusion processes and all values of $\alpha$. This result strongly resonates with previous findings on $1/f$ noise, $1/f$ noise with different levels of non-Gaussianity, and binomial multiplicative cascades \cite{kelty2022fractal,kelty2023multifractal,mangalam2022ergodic}.

\subsection{Multifractal spectrum distinguish different anomalous diffusion processes}

\begin{figure*}
\includegraphics{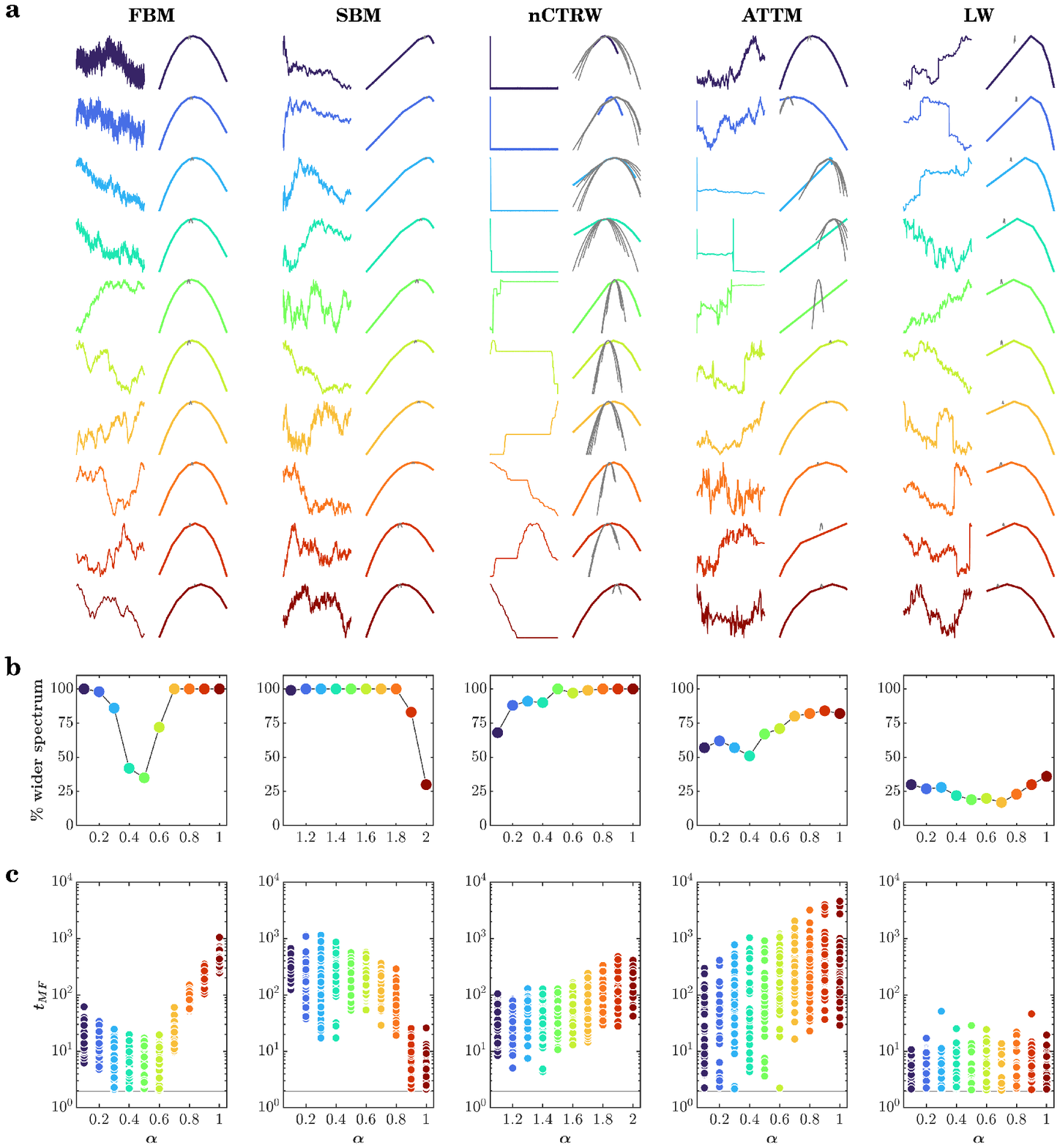}
\caption{Multifractal formalism appears to be as a nonlinear analytical method that unifies several disparate non-ergodic anomalous diffusion processes into a common framework of multiplicative cascades. (a) Multifractal spectrum for the five types of anomalous diffusion processes—FBM, SBM, nCTRW, ATTM, and LW. The anomalous exponent $\alpha$. $\alpha$ ranges from 0.1 to 1 for FBM, SBM, nCTRW, and ATTM and from 1.1 to 2 for LW, with an increment of 0.1 from dark blue to dark red. The thick colored curve in each plot indicate the multifractal spectrum fir the original series, and the thin grey curves in each plot indicate the multifractal spectrum for a sample of five corresponding IAAFT surrogates. The axes have been stretched to match the minimum and maximum values of ($\alpha (q)$,$f(q)$). (b) percentage of simulated trajectories with the wider-than-surrogates spectrum, i.e., $t_{MF} > 1.96$. (c) Multifractal nonlinearity $t_{MF}$ for processes with the wider-than-surrogates spectrum.}
\label{fig:f10}
\end{figure*}

Fig.~\ref{fig:f10}a shows the multifractal spectrum for the five processes—FBM, SBM, nCTRW, ATTM, and LW and different values of the anomalous exponent $\alpha$. FBM shows highly symmetric spectra with initially fleeting and then increasing difference between the original spectrum and those of the corresponding IAAFT surrogates for larger $\alpha$. This trend is confirmed by an initially fleeting and then increasing percentage of FBM trajectories with the wider-than-surrogate spectrum and an initially fleeting and then increasing $t_{MF}$ values with an increase in $\alpha$ (Figs.~\ref{fig:f10}b and \ref{fig:f10}c).

SBM shows highly asymmetric multifractal spectra, with the completely missing right half of the spectrum and with the spectra becoming more asymmetric with an increase in $\alpha $ and resembling the spectra for FBM at $\alpha = 1$—the particular case in which SBM is reduced to awGn (Fig.~\ref{fig:f10}a). This asymmetry reflects the putative effects of large $q$ moments, which exacerbate the effects of smaller fluctuations in the estimation of the multifractal spectrum. The difference between the original spectrum and those of the corresponding IAAFT surrogates is more prominent for smaller values of $\alpha$, a trend which was confirmed by the reduction in $t_{MF}$ values with an increase in $\alpha$ (Fig.~\ref{fig:f10}c).

nCTRW shows asymmetric multifractal spectra throughout but a more leftward skewed spectrum for $\alpha \rightarrow 1$ (Fig.~\ref{fig:f10}a). Again, this asymmetry reflects the putative effects of small $q$ moments, which exacerbate the effects of larger fluctuations in the estimation of the multifractal spectrum. However, despite these fluctuations in asymmetry of the multifractal spectrum, the difference between the original spectrum and those of the corresponding IAAFT surrogates increase sharply with $\alpha$, a trend confirmed by the increase in the percentage of nCTRW trajectories with the wider-than-surrogate spectrum and an increase in $t_{MF}$ values with an increase in $\alpha$ (Figs.~\ref{fig:f10}b and \ref{fig:f10}c).

ATTM shows highly asymmetric and distorted multifractal spectra, but the skew direction does not depend on $\alpha$ in a principled way (Fig.~\ref{fig:f10}a). Nonetheless, the difference between the original spectrum and those of the corresponding IAAFT surrogates increased sharply with $\alpha$. While the percentage of ATTM trajectories with the wider-than-surrogate spectrum increase with $\alpha$ (Fig.~\ref{fig:f10}b), $t_{MF}$ values show only marginal increase with $\alpha$ (Fig.~\ref{fig:f10}c).

LW shows highly asymmetric multifractal spectra, almost with missing left half of the spectrum when $\alpha \rightarrow 1$ (Fig.~\ref{fig:f10}a). This asymmetry reflects the putative effects of small $q$ moments, which exacerbate the effects of larger fluctuations in the estimation of the multifractal spectrum. The spectra of the original trajectories and the corresponding IAAFT surrogates also do not differ in a principled manner. This ambiguity is evident in the low percentage of LW trajectories with wider-than-surrogate spectrum ($< 50\;\text{out of}\;100$; Fig.~\ref{fig:f10}b) and comparable $t_{MF}$ values across all $\alpha$ (Fig.~\ref{fig:f10}c).

In short, the five anomalous diffusion processes all show multifractal evidence of nonlinearities, albeit minor differences in the percentage of trajectories showing multifractal evidence and the strength of evidence. Notably, the shape of the multifractal spectrum reflected the respective generative mechanism---e.g., symmetric spectra for FBM, left slewed spectra for SBM and right skewed spectra for LW reflecting the fact that smaller and larger fluctuations, respectively, characterize these two processes. While more detailed interpretations warrant further investigations, it is evident that multifractal analysis can diagnose certain differences among these anomalous diffusion processes.

\section{Discussion}

This work explores preliminary steps towards a unified framework grounded in the multifractal formalism \cite{ihlen2012introduction,kelty2013tutorial,kelty2022multifractaltest} for restoring ergodicity to a description of anomalous diffusion processes \cite{ritschel2021universality,vinod2022nonergodicity,wang2022restoring}. We used synthetic data representing various anomalous diffusion processes for a wide range of anomalous exponent $\alpha$, both ergodic and non-ergodic, approximated by five disparate mathematical models: FBM, ergodic; SBM, weakly non-ergodic; CTRW, weakly non-ergodic; ATTM, weakly non-ergodic; and LW, ultra-weakly non-ergodic. We show that $\varA{TA-MSD}$ and $\varA{MSD}$-related linear descriptors such as $SD$, $CV$, and $RMS$ break ergodicity. In contrast, time series of descriptors addressing sequential structure and its potential nonlinearity: multifractality, and, to some extent, fractality, change in a time-independent way and are ergodic descriptors insensitive to the weak ergodicity breaking of the process. Thus, these descriptors return the same information for any kind of diffusion process and the anomalous exponent $\alpha$. Further analysis revealed that these findings directly followed the multiplicative cascades underlying these diffusion processes, as the shape and symmetry of the multifractal spectrum—and those of the corresponding surrogate series—differentiated these processes. Thus, the statistical descriptors analyzed here provide very different, complementary information to other statistical descriptors. Two particular points bear emphasis here. First, because multifractal descriptors of anomalous diffusion remain ergodic, they can be submitted to linear causal modeling. Second, this capacity to describe non-ergodic anomalous diffusion processes in ergodic terms offers the possibility that multifractal modeling could unify these processes into a common framework.

Multifractal formalisms can serve as the desired analytical framework for linear causal modeling of anomalous diffusion processes. Whereas $\varA{TA-MSD}$ and $\varA{TA-MSD}$-related linear descriptors like $SD$, $CV$, and $RMS$ that are typically submitted to linear causal models \cite{kian2017relationship,vink2017eeg,vink2020eeg} break ergodicity \cite{mangalam2021point,mangalam2022ergodic,kelty2022fractal}. In contrast, multifractal descriptors remain ergodic and hence, offer a reliable and stable set of causal predictors \cite{bloomfield2021perceiving,booth2018expectations,carver2017multifractal,dixon2012multifractal,jacobson2021multifractality,kelty2014interwoven,kelty2021multifractal,mangalam2020bodywide,mangalam2020global,mangalam2020multifractal,mangalam2020multiplicative,wallot2018interaction}. Although $\textrm{MSD}$ remains prevalent in measuring active matter, our present results resonate with a growing interest in multifractal modeling in many of these active-matter fields, e.g., bio-molecules moving within cells \cite{cardenas2012dynamics,chaieb2008wrinkling,rezania2021multifractality,wawrzkiewicz2020multifractal}, animals foraging in the wild \cite{gutierrez2015neural,ikeda2020c,schmitt2001multifractal,seuront2014anomalous}, and the emergence of collective dynamics such as swarming and milling \cite{balaban2018quantifying,carver2017multifractal,koorehdavoudi2017multi}, have begun to embrace multifractal formalisms. We hope that the current findings might emphasize the importance of these approaches.

The ergodicity of multifractal descriptors allows the possibility that cascade dynamics constitute a statistically testable causal framework that may explain these disparate anomalous-diffusion regimes. The current findings underscore that multifractal structure is not merely an abstract side-effect, nor is it a nuisance to be collapsed into the noise terms to avoid cluttering the lower-dimensional aspects of our generative models \cite{chen1997long,slifkin2020trajectory}. The accumulating evidence of multifractal structure and its relevance for describing and predicting structural change has implicated a causal role in cascading dynamics \cite{kelty2021multifractal,lovejoy2018weather,mandelbrot1974intermittent}. Indeed, the high-dimensional aspect of cascade dynamics sometimes raises new and unfamiliar questions about the relationship between causality and low-dimensional determinism. Indeed, we may feel most confident explaining when we have reduced our model systems to a minimal set of control parameters. That confidence may lead us to take for granted that the low-dimensional constraint needed for deterministic modeling is required to model and explain causal relationships. Our commitment to low-dimensional causation is so strong that scholars will even reason that low-dimensionality and so determinism is a matter of observer’s knowledge: e.g., the suggestion that, if only we knew how the system works, then we might no doubt see that causation is low-dimensional after all \cite{karhausen2001commentary}. However, the mutually fostering growth of multifractal estimation and cascade-dynamical modeling has strengthened the possibility that causation may not need low dimensionality. Philosophical, logical, and empirical approaches have all begun to point to growing comfort and fluency with the concepts of stochastic causation \cite{bahar2006increased,lovejoy2018weather,sanabria2020internal,shahar2007estimating,vijg2020loss}, and even stochastic-deterministic blends that reflect the cascade-like dynamics across multiple scales \cite{van2015active}.

The observed variety of multifractal spectrum across the five simulated anomalous diffusion processes suggests that multifractal formalisms could also aid in time-series characterization and clustering. The rate of generation of time series is exponentially increasing in all areas of physical and life sciences, and the production of ad-hoc analytical tools is accompanying this growth \cite{cliff2022unifying,fulcher2013highly,fulcher2017hctsa}. Many of these attempts use Bayesian \cite{krog2017bayesian,krog2018bayesian,park2021bayesian,thapa2018bayesian,thapa2022bayesian} and ML approaches \cite{bo2019measurement,cichos2020machine,granik2019single,janczura2020classification,munoz2020single,munoz2021objective}, and even unsupervised \cite{pineda2022geometric,gajowczyk2021detection,gentili2021characterization,kowalek2022boosting,munoz2021unsupervised,seckler2022bayesian} to detect specific anomalous diffusion processes and the underlying mathematical model, especially deviation from pure Brownian behavior in terms of the anomalous exponent. However, these attempts still lack the accuracy, sensitivity, and specificity necessary, say, for understanding how diffusion properties change over time due to environmental heterogeneity (e.g., patches with different viscosity on a cellular membrane), time-varying properties of the observable (e.g., different activation states of a molecular motor). This limitation may reflect that these attempts typically rely on manual or automatic extraction of features that may not have to do with plausible generating mechanisms \cite{kowalek2019classification,loch2020impact}. Including multifractal descriptors with $\varA{MSD}$-related linear descriptors might improve the ML-powered characterization and clustering of anomalous diffusion processes.

Interdependent fluctuations can cause interactions across a wide range of spatiotemporal scales, altering the context for subsequent fluctuations. Cascade instabilities, e.g., can produce turbulent structures, which are complex flows in which once-parallel currents collapse or explode into a dizzying, possibly limitless variety of vortices and eddies, with intermittent swelling and ebbing throughout space and time \cite{lovejoy1998diffusion,mandelbrot1974intermittent,shlesinger1987levy}. Indeed, anomalous diffusion and Lévy walks distinguish active from inertial turbulence \cite{mukherjee2021anomalous}. Our results indicate that the various diffusion coefficients are thoroughly interconnected with the specific geometries of fluctuations constituting the measured series. The connection between multifractality and various models of anomalous diffusion is also being noticed both theoretically \cite{afanasiev1991chaotic,chen2010anomalous,de2004anomalous,gmachowski2015fractal,lim2002self} and empirically \cite{bickel1999simple,lovejoy2018weather,menu2020anomalous,schmitt2001multifractal,seuront2004random,seuront2014anomalous,sharifi2012investigation}. Multifractal formalisms and anomalous-diffusion processes thus appear to be entwined in a vibrant, expanding, far-reaching, and synergistic relationship originating from the out-of-equilibrium character, lack of detailed balance, and of time-reversal symmetry, multiscale nature, nonlinearity and multi-body interactions that typify living and evolving systems \cite{shaebani2020computational}. Future investigations could further explore the relationship between the various features of the multifractal spectrum and anomalous diffusion.

\appendix

\section{Theoretical models}

\subsection{Fractional Brownian motion}

\noindent{}In fractional Brownian motion (FBM), $x(t)$ is a Gaussian process with stationary increments, it is symmetric, $\langle x(t) \rangle = 0$, and importantly its $\varA{EA-MSD}$ scales as $\langle x(t) \rangle =2K_{H}t^{2H}$, where $H$ is the Hurst exponent and is related to the anomalous exponent $\alpha$ as $H=\alpha/2$ \cite{jeon2010fractional,mandelbrot1968fractional}. The two-time correlation for FBM is $\langle x(t_{1}) x(t_{2}) \rangle = K_{H}(t_{1}^{2H}+t_{2}^{2H}-|t_{1}-t{2}|^{2H})$. FBM can also be defined as a process that arises from a generalized Langevin equation with non-white noise (or fractional Gaussian noise, $fGn$). The $fGn$ has a standard normal distribution with zero mean and power-law correlations:
\begin{multline*}
  < \xi_{fGn}(t_{1}) \xi_{fGn}(t_{2}) > = 2K_{H}H(2H-1)|t_{1}-t{2}|^{2H-2} + \\ 4K_{H}H|t_{1}-t{2}|^{2H-1}\delta(t_{1}-t{2}). \tag{A1}\label{A1}
\end{multline*}

The FBM features two regimes: one in which the noise is positively correlated ($1/2<H<1$, i.e., $1<\alpha<2$, superdiffusive) and the other in which the noise is negatively correlated ($0<H<2$, i.e., $0<\alpha<1$, subdiffusive). For $H=1/2$ ($\alpha=1$) the noise is uncorrelated, hence the FBM converges to Brownian motion.

Various numerical approaches have been proposed to solve the FBM generalized Langevin equation. We use the method described by Bardet et al. \cite{bardet2003generators} via the Matlab function \texttt{wfbm()}. Details about the numerical implementations can be found in the associated
reference.

\subsection{Scaled Brownian motion}

\noindent{}The scaled Brownian motion (SBM) is a process described by the Langevin equation with a time-dependent diffusivity
\begin{equation*}
  \frac{dx(t)}{dt} = \sqrt{2Kt} \xi(t), \tag{A2}\label{A2}
\end{equation*}
where $\xi = 1$ is white Gaussian noise \cite{munoz2022stochastic}. In the case when $K(t)$ has a power-law dependence on to $t$ such that $K(t) = \alpha K_{\alpha} t^{\alpha - 1}$, $\varA{EA-MSD}$ follows $<x^{2}(t)>_{N} = K_{\alpha}t^{\alpha}$ with $K(t) = \Gamma(1 + \alpha)K_{\alpha}$. The numerical implementation of SBM is presented in \textit{Algorithm 1}.\\

\noindent{}\textbf{Algorithm 1:} Generate SBM trajectory

\textbf{Input:}

\setlength{\parindent}{20pt}
length of the trajectory $T$

anomalous exponent $\alpha$
\setlength{\parindent}{10pt}

\textbf{Define:}

\setlength{\parindent}{20pt}
\texttt{erfcinv}($\vec a$) $\rightarrow$ Inverse complementary error function of $\vec a$

$U(L) \rightarrow$ returns $L$ uniform random numbers $\in[0,1]$
\setlength{\parindent}{0pt}

\textbf{Calculate:}

\setlength{\parindent}{20pt}
$\vec {\Delta x} \leftarrow (1^{\alpha}, 2^{\alpha}, ..., T^{\alpha})-(0^{\alpha}, 1^{\alpha}, ..., (T - 1)^{\alpha})$

$\vec {\Delta x} \leftarrow 2 \sqrt(2) UL \vec {\Delta x}$

$\vec x \leftarrow$ cumsum ($ \vec {\Delta x}$)
\setlength{\parindent}{0pt}

\textbf{Return:} \textbf{$\vec x$}

\subsection{Noisy continuous-time random walk.}

\noindent{}The continuous-time random walk (CTRW) is a family of random walks with arbitrary displacement density for which the waiting time, i.e., the time between subsequent steps, is a stochastic variable \cite{scher1975anomalous}. We considered a specific case of CTRW with waiting times following a power-law distribution $\psi(t) = t^{-\sigma}$ and displacements following a Gaussian distribution with variance $D$ and zero mean. In such case, the anomalous exponent is $\alpha = \sigma-1$ ($\varA{EA-MSD} = \langle x(t)^{2} \rangle \propto t^{\alpha}$). To obtain noisy CTRW (nCTRW) \cite{jeon2013noisy}, white Gaussian noise with 0 mean and standard deviation equal to the standard deviation of the corresponding CTRW fluctuation series was added to each CTRW series. Since the waiting times follow a power-law distribution, for $\sigma = 2$, $\varA{EA-MSD}$ features Brownian motion with logarithmic corrections \cite{klafter2011first}.

\setlength{\parindent}{10pt}The numerical implementation of CTRW is presented in \textbf{Algorithm 2}. Notice that the variable $\tau$ stands for the total time at $i$-th iteration. The output vector $\vec x$ corresponds to the position of the particle at the irregular times given by $\vec t$.\\

\noindent{}\textbf{Algorithm 2:} Generate CTRW trajectory

\textbf{Input:}

\setlength{\parindent}{20pt}
length of the trajectory $T$

anomalous exponent $\alpha$

diffusion coefficient $D$
\setlength{\parindent}{0pt}

\textbf{Define:}

\setlength{\parindent}{20pt}
$\vec x \rightarrow$ empty vector

$\vec t \rightarrow$ empty vector

$N(\mu,S) \rightarrow$ Gaussian random number generator with mean $\mu$ and standard deviation $s$
\setlength{\parindent}{0pt}

$i = 0$; $\tau = 0$

\textbf{While} $\tau < T$ \textbf{do}

\setlength{\parindent}{15pt}
$t_{i}$ sample randomly from $\psi(t) \sim t^{- \sigma}$

$x_{i} \leftarrow x_{i - 1} + N(0,\sqrt{D})$

$\tau \leftarrow \tau + t_{i}$

$i \leftarrow i + 1$
\setlength{\parindent}{0pt}

\textbf{end while}

\textbf{Return:} \textbf{$\vec x$}, \textbf{$\vec t$}

\subsection{Annealed transient time motion}

\noindent{}The annealed transient time motion (ATTM) implements the motion of a Brownian particle with time-dependent diffusivity \cite{massignan2014nonergodic}. The observable performs Brownian motion for a random time $t_{1}$ with a random diffusion coefficient $D_{1}$, then for $t_{2}$ with $D_{2}$, and so on. The diffusion coefficients follow a distribution such that $P(D) = D^{\sigma - 1}$ with $\sigma > 0$ as $D \to 0$ and that decays rapidly for large $D$. If the random times $t$ are sampled from a distribution with expected value $E[t|D] = D^{-\gamma}$, with $\sigma < \gamma < \sigma+1$, the anomalous exponent is $\alpha = \sigma/\gamma$. Here, we consider that the distribution is a delta function, $P_{t}(t|D) = \delta(1 - D^{-\gamma})$. Hence, the time $t_{i}$ in which the observable performs Brownian motion with a random diffusion coefficient $D_{i}$ is $t_{i} = D_{i}^{-\gamma}$, with $D_{i}$ extracted from the distribution described above.

\setlength{\parindent}{10pt}The numerical implementation of ATTM is presented in \textit{Algorithm 3}. Note that, in contrast to nCTRW and LW, now the only output is $\vec x$ because the trajectory is already produced at regular time intervals.\\

\noindent{}\textbf{Algorithm 3:} Generate ATTM trajectory

\textbf{Input:}

\setlength{\parindent}{20pt}
length of the trajectory $T$

anomalous exponent $\alpha$

sampling time $\Delta t$
\setlength{\parindent}{0pt}

\textbf{Define:}

\setlength{\parindent}{20pt}
\textbf{While} $\sigma > \gamma$ and $\gamma > \sigma + 1$ \textbf{do}

\setlength{\parindent}{30pt}
$\sigma \leftarrow$ uniform random number $\in(0,3]$

$\gamma = \sigma / \alpha$
\setlength{\parindent}{20pt}

\textbf{end while}

BM($D$,$t$,$\Delta t$) $\rightarrow$ generates a Brownian motion trajectory of length $t$ with diffusion coefficient $D$, sampled at time intervals $\Delta t$
\setlength{\parindent}{0pt}

\textbf{While} $\tau < T$ \textbf{do}

\setlength{\parindent}{15pt}
$D_{i} \leftarrow$ sample randomly from $P(D)D^{\sigma - 1}$

$t_{i} \leftarrow D_{i}^{- \gamma}$

number of steps $N_{i} = round(t_{i}/ \Delta t)$

$x_{1}, ..., x_{i + N_{i}} \leftarrow$ BM($D$,$t$,$\Delta t$)

$i \leftarrow i + N_{i} + 1$

$\tau = \tau + N_{i} \Delta t$
\setlength{\parindent}{0pt}

\textbf{end while}

\textbf{Return:} $\vec x$

\subsection{Lévy walk}

\noindent{}The Lévy walk (LW) is a particular case of superdiffusive CTRW. Like subdiffusive CTRW, the flight time, i.e., the time between steps, for LW is irregular \cite{klafter1994levy}, but, in contrast to subdiffusive CTRW, the distribution of displacements for LW is not Gaussian. We considered the case in which the flight times follows the distribution $\psi(t) = t^{- \sigma - 1}$. At each step, the displacement is $\Delta x$ and the step length is $|\Delta x|$. The displacements are correlated with the flight times such that the probability to move a step $\Delta x$ at time $t$ and stop at the new position to wait for a new random event to happen is $\psi(\Delta x,t) = \frac{1}{2} \delta(|\Delta x|-vt) \psi(t)$, where $v$ is the velocity. The anomalous exponent is given by
\begin{equation*}
\mathrm{EB}(x(t)) =
  \begin{cases}
  2 & \text{if } 0 < \sigma < 1 \\
  3 - \sigma & \text{if } 1 < \sigma < 2.
  \end{cases}
  \tag{A3}\label{eq:A3}
\end{equation*}

\setlength{\parindent}{10pt}The numerical implementation of LW is presented in \textit{Algorithm 4}. Notice that we use a random number $r$, which can take values 0 or 1, to decide in which sense the step is performed. The output vectors $\vec x$ represent irregularly sampled positions and times.\\

\noindent{}\textbf{Algorithm 4:} Generate LW trajectory

\textbf{Input:}

\setlength{\parindent}{20pt}
length of the trajectory $T$

anomalous exponent $\alpha$
\setlength{\parindent}{0pt}

\textbf{Define:}

\setlength{\parindent}{20pt}
$\vec x \rightarrow$ empty vector

$\vec t \rightarrow$ empty vector

$v \rightarrow$ random number $\in (0,10]$
\setlength{\parindent}{0pt}

$i=0$

\textbf{While} $\tau < T$ \textbf{do}

\setlength{\parindent}{15pt}
$t_{i} \leftarrow$ sample randomly from $\psi(t) \sim t_{- \sigma - 1}$

$x_{i} (-1)^{r}vt_{i}$, where random $r$ is 0 or 1 with equal probability.

$\tau \leftarrow \tau + t_{i}$

$i \leftarrow i + 1$
\setlength{\parindent}{0pt}

\textbf{end while}

\textbf{Return:} $\vec x, \vec t$

\bibliography{apssamp}

\end{document}